\documentclass[useAMS,usenatbib,a4paper]{mn2e}
\usepackage{psfig}
\usepackage{xspace}
\usepackage{graphicx}
\usepackage{amssymb}
\usepackage{amsmath}
\usepackage{longtable}

\graphicspath{{sw-cfhtls-figs/}{./}}

\newcommand{\be}{\begin{equation}}
\newcommand{\ee}{\end{equation}}
\newcommand{\bea}{\begin{eqnarray}}
\newcommand{\eea}{\end{eqnarray}}

\def\Sref#1{Section~\ref{#1}\xspace}
\def\Fref#1{Figure~\ref{#1}\xspace}
\def\Tref#1{Table~\ref{#1}\xspace}
\def\Eref#1{Equation~\ref{#1}\xspace}
\def\Aref#1{Appendix~\ref{#1}\xspace}

\newcommand{\PaperOne}{Paper~I\xspace}
\newcommand{\PaperTwo}{Paper~II\xspace}

\newcommand{\StageOne}{Stage~1\xspace}
\newcommand{\StageTwo}{Stage~2\xspace}

\newcommand{\kms}{\ifmmode  \,\rm km\,s^{-1} \else $\,\rm km\,s^{-1}  $
\fi }
\newcommand{\kpc}{\ifmmode  {\rm kpc}  \else ${\rm  kpc}$ \fi  }
\newcommand{\pc}{\ifmmode  {\rm pc}  \else ${\rm pc}$ \fi  }
\newcommand{\Msun}{\ifmmode {\rm M_{\odot}} \else ${\rm M_{\odot}}$ \fi}
\newcommand{\Zsun}{\ifmmode {\rm Z_{\odot}} \else ${\rm Z_{\odot}}$ \fi}
\newcommand{\yr}{\ifmmode yr^{-1} \else $yr^{-1}$ \fi}
\newcommand{\hMsun}{\ifmmode h^{-1}\,\rm M_{\odot} \else $h^{-1}\,\rm
M_{\odot}$ \fi}

\def\sw{{\small\sc Space\,Warps}\xspace}
\def\SW{{\sc Space\,Warps}\xspace}
\def\Talk{{\small\sc Talk}\xspace}

\def\cfhtls{{CFHTLS}\xspace}

\def\gravlens{{\sc gravlens}\xspace}
\def\sextractor{{\sc SExtractor}\xspace}

\def\humvi{{\sc HumVI}\xspace}
\def\af{{\sc ArcFinder}\xspace}

\def\rf{{\sc RingFinder}\xspace}

\def\GZ{{\sc Galaxy\,Zoo}\xspace}

\def\CM{\mathcal{M}}
\def\PL{\CM_{LL}}
\def\PD{\CM_{NN}}

\usepackage[usenames]{color}

\def\oxford{Dept.\ of Physics, University of Oxford, Keble Road, Oxford, OX1 3RH, UK}

\def\kipac{Kavli Institute for Particle Astrophysics and Cosmology, Stanford University, 452 Lomita Mall, Stanford, CA 94035, USA}
\def\ipmu{Kavli IPMU (WPI), UTIAS, The University of Tokyo, Kashiwa, Chiba 277-8583, Japan}
\def\zooniverse{Zooniverse, c/o Astrophysics Department, University of Oxford, Oxford OX1 3RH, UK}
\def\adler{Adler Planetarium, Chicago, IL, USA}

\def\zurich{Department of Physics, University of Zurich, Winterthurerstrasse 190, 8057 Zurich, Switzerland}
\def\paris{Institut d’Astrophysique de Paris, UMR7095 CNRS – Universit\'e Pierre et Marie Curie, 98bis bd Arago, 75014 Paris, France}
\def\icg{Institute of Cosmology and Gravitation, University of Portsmouth, Dennis Sciama Building, Portsmouth P01 3FX, UK}

\def\amemail{\tt anupreeta.more@ipmu.jp}

\title[New Gravitational Lens Candidates from CFHTLS]
{\SW: II. New Gravitational Lens Candidates from the CFHTLS Discovered
through Citizen Science}

\author[More et al.]{%

   \newauthor{%
    Anupreeta~More,$^{1}$\thanks{\amemail}
    Aprajita~Verma,$^{2}$
    Philip~J.~Marshall,$^{2,3}$
    Surhud~More,$^{1}$
    }
   \newauthor{%
    Elisabeth~Baeten,$^{4}$
    Julianne~Wilcox,$^{4}$
    Christine~Macmillan,$^{4}$
    Claude~Cornen,$^{4}$
   }
   \newauthor{%
    Amit~Kapadia,$^{5}$
    Michael~Parrish,$^{5}$
    Chris~Snyder,$^{5}$
    Christopher~P.~Davis,$^{3}$
    }
   \newauthor{%
    Raphael~Gavazzi,$^{6}$
    Chris~J.~Lintott,$^{2}$
    Robert~Simpson,$^{2}$
    David~Miller,$^{4}$
    Arfon~M.~Smith,$^{4}$
    }
   \newauthor{%
    Edward~Paget,$^{4}$
    Prasenjit~Saha,$^{7}$
    Rafael~K\"ung,$^{7}$
    Thomas~E.~Collett$^{8}$
    }
\medskip\\
$^1$\ipmu\\
$^2$\oxford\\
$^3$\kipac\\
$^4$\zooniverse\\
$^5$\adler\\
$^6$\paris\\
$^7$\zurich\\
$^8$\icg\\

}

\begin{document}

\date{to be submitted to MNRAS}
\pagerange{\pageref{firstpage}--\pageref{lastpage}}\pubyear{2013}

\maketitle

\label{firstpage}


\begin{abstract}
We report the discovery of 29 promising (and 59 total) new lens
candidates from the CFHT Legacy Survey (\cfhtls) based on about 11
million classifications performed by citizen scientists as part of the
first \sw lens search.  The goal of the blind lens search was to
identify lens candidates missed by robots (the \rf on galaxy scales and
\af on group/cluster scales) which had been previously used to mine the
\cfhtls for lenses.  We compare some properties of the samples detected
by these algorithms to the \sw sample and find them to be broadly
similar.  The image separation distribution calculated from the \sw
sample shows that previous constraints on the average density profile of
lens galaxies are robust. \sw recovers about 65\% of known lenses, while
the new candidates show a richer variety compared to those found by the
two robots. This detection rate could be increased to 80\% by only using
classifications performed by expert volunteers (albeit at the cost of a
lower purity), indicating that the training and
performance calibration of the citizen scientists is very important for
the success of \sw. In this work we present the SIMCT pipeline, used for
generating {\it in situ} a sample of realistic  simulated lensed images.
This training sample, along with the false positives identified during
the search, has a legacy value for testing future lens finding
algorithms. We make the pipeline and the training set publicly
available.
\end{abstract}


\setcounter{footnote}{1}


\section{Introduction}
\label{sec:intro}

The last few decades have seen a rise in the discoveries of strong
gravitational lenses owing to the plethora of interesting applications
lenses have in astrophysics and cosmology. Strong lenses are routinely
used to probe the dark matter distribution from galaxy
\citep[e.g.][]{Koopmans2006,Barnabe2009,Leier2011,Sonnenfeld2015} to group and
cluster scales
\citep[e.g.][]{Limousin2008,Zitrin2011,Oguri2012,More2012,Newman2013}, to
study distant young galaxies by using the lensing magnification as a
natural telescope \citep[e.g.][]{Zitrin2009,Zheng2012,Whitaker2014}, to
test the cosmological model by constraining cosmological parameters such
as the Hubble constant and the dark energy equation of state
\citep[e.g.][]{Suyu2010,Collett2012,Collett2014,Sereno2014}, and many more. Strong
lenses are rare, because a foreground massive object needs to be
sufficiently aligned with a distant background source to produce
multiple images. Nevertheless, systematic lens searches have led to the
discovery of over 500 lenses to
date.\footnote{\texttt{http://admin.masterlens.org/index.php}}

The search for gravitational lenses is a needle-in-a-haystack problem.
Several automated lens finding algorithms have
been developed so far
\citep[e.g.][]{Lenzen2004,Alard2006,Seidel2007,More2012,Brault2014,Gavazzi2014},
but they can not
simultaneously capture the myriad types of lenses that are known to
exist. For example, the lensed images of background galaxies show
variety in their surface brightness distributions, colours, light
profiles, shapes, structures and angular image separations.  Moreover,
many lensed images appear similar to features found commonly in galaxies
(such as spiral arms) or to artefacts in astronomical images (scattered
light around stars).  Almost all lens finding algorithms find it
difficult to distinguish these from the real lenses and thus suffer
from a high rate of false positive detections.  To mitigate this problem,
algorithms are often restricted to detect a very narrow class of lens
systems. However, even after such restrictions, robotic lens searches
have to always rely on visual screening to produce a sample of plausible
lens candidates.

Recognising patterns is one of the
strengths of the human brain. Humans are also capable of dealing with
multi-tiered complex web of questions before arriving at a conclusion, a process
which may
not be always possible to automate. The algorithm by which our brains process a
task is extremely malleable, self-learning and self-evolving. Therefore it has
a huge potential for the discovery of exotic objects which do not quite fit a
set criteria, but are still very likely to be objects of interest. The lens
finding algorithms are not yet advanced enough to produce better performance
than visual classifications. Consequently, as we enter the era of
large area imaging surveys spanning thousands of square degrees,
the participation of a large community of
volunteers to help with the visual identification of lenses would be very
beneficial for the lensing community.  Now seems
the perfect time to investigate the potential of citizen science.

\GZ, one of the most successful citizen science projects in
astronomy, addressed
the problem of how to  classify large numbers of galaxies by their
morphology \citep{Lintott2008}.
From these early results to several new unexpected and
interesting discoveries, such as that of green pea galaxies
\citep{Cardamone2009,Jaskot2013} and Hanny's Voorwerp
\citep{Lintott2009,Keel2012}, \GZ has been able to start to realize the potential
of citizen scientists.  Since then, both astronomy and non-astronomy projects
have been launched under the citizen science web portal Zooniverse
(\texttt{http://zooniverse.org}). The task of finding gravitational lenses is
significantly challenging, given that the lens systems show such
complexity and that they are rare. To add to the challenge, not many
citizen scientists are expected to be aware of the phenomenon of
gravitational lensing, and the resulting characteristic image configurations.  With
these significant challenges at hand, we designed the \sw project to
enable the discovery of lenses through citizen science (learning from
previous experience in serendipitous identification of lens candidates in \GZ).
In a companion
paper \citep[][hereafter \PaperOne]{Marshall2015}, we describe the design of
\sw and how the entire system functions as a discovery service.  In this
paper (\PaperTwo), we describe our first lens search using data
from the Canada-France-Hawaii Telescope Legacy Survey
(\cfhtls\footnote{\texttt{http://www.cfht.hawaii.edu/Science/CFHTLS/}}).
In \citet{Kung2015}, we describe the design of a collaborative
mass modelling tool that can be used by citizen scientists.

This paper is organised as follows. In
\Sref{sec:data}, we introduce the \cfhtls imaging data and the
previously published lens samples from the \cfhtls. We generated a training sample,
consisting of simulated lenses, duds and impostors, in order to
aid the \sw volunteers in the process of finding lenses. We give details of this
training sample in \Sref{sec:ts} and \Sref{sec:dfp}. In \Sref{sec:swap},
we briefly describe how the classifications of images from the volunteers are
turned into a catalog of plausible candidates (for further details, see
\PaperOne). In \Sref{sec:results}, we present the new lens candidates from \sw and
compare it to the lens samples produced by past robotic searches of
the \cfhtls. Next, we discuss what kind of lenses are detected
or missed by the algorithms and \sw in \Sref{sec:discuss}.
Our conclusions are given in \Sref{sec:conclude}.


\section{Data}
\label{sec:data}

\subsection{The CFHT Legacy Survey}
\label{sec:data:cfhtls}

The \cfhtls is a photometric survey in five optical bands
($u^*g'r'i'z'$) carried out with the wide-field imager MegaPrime which
has a 1~deg$^2$ field-of-view and a pixel size of 0.186\arcsec
\citep{Gwyn2012}. The \cfhtls WIDE covers a total
non-overlapping area of 160~deg$^2$ on the sky and consists of four
fields W1, W2, W3 and W4. The field W1 has the largest sky coverage of
63.7~deg$^2$. The fields W2 and W4 have similar sky coverages of
22.6~deg$^2$ and 23.3~deg$^2$, respectively\footnote{These numbers are
estimated from
\texttt{http://terapix.iap.fr/cplt/T0007/doc/T0007-doc.pdf}}.  The field
W3 has a sky coverage of 44.2~deg$^2$ and is more than twice as large as
W2 and W4.

The \cfhtls imaging is very homogeneous and has good image quality. Most
of the lensed arcs are much brighter in the $g$ band, so deep imaging
in this band is desirable. The limiting magnitude is 25.47 for the $g$
band which goes the deepest among all of the five bands. The mean seeing
in the $g$ band is 0.78\arcsec. The zero point to convert flux to AB
magnitude for all bands is 30. These characteristics make \cfhtls ideal
to do visual inspection for finding lenses.  We use the stacked images
from the final T0007 release taken from the Terapix
website\footnote{\texttt{http://terapix.iap.fr/cplt/T0007/doc/T0007-doc.pdf}}
for this work.

We note that the \cfhtls is a niche survey with a unique combination of wide
imaging with deep sensitivity. It is a precursor to the ongoing wide imaging
surveys such as the Dark Energy Survey (DES), Kilo Degree Survey (KiDS) and the
Hyper Suprime-Cam (HSC) survey and other planned future surveys such as the
Large Synoptic Survey Telescope (LSST) survey.  The search for lenses with \sw
in the \cfhtls is an important step to learn lessons and prepare for lens
searches in these larger imaging surveys.


\subsection{Previously published lens samples from the \cfhtls}
\label{sec:data:kls}

The \cfhtls has been searched for lenses using various lens finding
methods and algorithms. Here, we give a brief summary of previously published lens
samples in the chronological order.

From the early release of the CFHTLS (T0002) covering 28~deg$^2$,
\citet{Cabanac2007} used an arc finding algorithm \citep{Alard2006} to
find arcs in galaxies, groups and clusters. They found about 40 lens
candidates with quality grades from low to high.

In the thesis dissertation of \citet{Thanjavur2009}, 9 promising and 2
low probability candidates were reported as having been discovered
serendipitously. These detections were made during the visual
inspection of the \cfhtls images as part of data reduction procedures
for the Weak Lensing survey \citep{Benjamin2007}.

\citet{Sygnet2010} carried out a search for edge-on disk galaxy lenses in the
\cfhtls WIDE. They identified galaxies, using \sextractor, which
had $18<i<21$ and inclination angle $<25\,^{\circ}$. After applying few more
selection criteria and visual inspection, they found about 3 promising
and a total of 18 lens candidates.

The \af \citep{More2012} was used for finding blue arc-like features in
the entire \cfhtls imaging without any pre-selection on the type of the
lensing object. This \af, an improved version of the algorithm by
\citet{Alard2006}, measures the second order moments of the flux
distribution in pixels within small regions to estimate the direction
and extent of local elongation of features. Pixels with high values of
elongation are connected to form an arc candidate. Finally, a set of
thresholds on arc properties such as the area, length, width, curvature
and surface brightness are used to select arc-like candidates. The
search was carried out in the $g$-band which is the most efficient
wavelength to find typical lensed features.  This sample, called SARCS,
has 55 promising and a total of 127 lens candidates which are selected
from both \cfhtls WIDE and DEEP fields. The SARCS sample consists of
some galaxy-scale candidates and mostly groups/cluster scale lens
candidates. This is because more massive systems produce arcs or lensed
images with large image separation from the lensing galaxy which are
easier to detect compared to the galaxy-scales. In the absence of a
large systematically followed up verified sample of candidates, we
choose the most promising 26 systems as our bona fide lens sample from
the \cfhtls WIDE. The total number of lens candidates in the \cfhtls
WIDE alone is 108.

In \citet{Elyiv2013}, the authors visually inspected a sample of 5500 optical
counterparts of X-ray point-like sources identified in the XMM-LSS
imaging of the \cfhtls W1 field. The goal was to find instances
of lensed quasars. Their sample consists of a total of 18 candidates,
of which 3 candidates were found to be promising.

\citet{Gavazzi2014} used their \rf code to find compact rings or arcs
around centres of isolated and massive early-type galaxies. \rf
subtracts the point spread function (PSF)-matched $i$-band images from the $g$-band images, and
looks for excess flux in the bluer $g$-band. An object detector measures
the properties of these residual blue features, and candidates which
meet the length-width ratio and tangential alignment criteria are then
visually inspected to form the final sample. \citet{Gavazzi2014}
pre-selected $\sim$638,000 targets as either photometrically-classified
early type galaxies, or objects selected to have red centres and blue
outer parts, from the T0006 CFHTLS data release catalogs. A total of
14370 galaxies were found to show detectable blue residuals, and 2524
were visually inspected, having passed the automatic feature selection
process. This led to a total of 330 lens candidates out of which 42 were deemed
good quality (\texttt{q\_flag} = 3) and 288  medium quality
(\texttt{q\_flag} = 2) candidates. In addition to the main well-defined
sample of \citet{Gavazzi2014}, a further 71 candidates were reported to
have been detected by earlier versions of the \rf, or from the \cfhtls DEEP.
From the main sample of ``\rf candidates,'' the SL2S team found, during
their follow-up campaign, 33 confirmed lenses \citep{Sonnenfeld2013a,Sonnenfeld2013b}.

The work by \citet{Maturi2014} used the arc finding code of
\citet{Seidel2007} and colour properties of typical arcs to optimize arc
detection. This new approach was tested on the CFHTLS-Archive-Research
Survey \citep[CARS,][]{Erben2009} which covers an area of 37~deg$^2$
only, and this entire image set was also visually inspected by the authors to
estimate the completeness and purity of their robotic search. They found 29
candidates with the robotic search alone and 41 candidates through pure
visual inspection---some of which were known from previous searches. Most
of these candidates are medium-low
probability\footnote{\texttt{http://www.ita.uni-heidelberg.de/$\sim$maturi/Public/arcs}}.

The \rf and the \af searches are the only searches that make use of a
lens finding algorithm and that have been run on the entire \cfhtls
imaging dataset.  Thus, we considered these to be our reference sample of known
lenses from robotic searches. For the purposes of transparency and to
help with the training, the volunteers participating in \sw-\cfhtls lens
search were made aware of these two known lens samples. Images
containing the systems from the \rf and the \af samples were labelled as
``known lens candidates'' in the \sw discussion forum,
\Talk,\footnote{\texttt{http://talk.spacewarps.org/}} where volunteers have the
opportunity to discuss their findings with fellow volunteers and the
science team. In this paper, we refer to the sample of 330 \rf and 108
\af lens candidates as the sample of ``known lens candidates'' and the
sample of confirmed (or most promising) 33 \rf and 26 \af as the sample
of ``known lenses''. Note that the ``known lens'' sample is a subset of
the ``known lens candidates'' sample. Also, note that the lens
candidates from the other papers listed above were not included in our
reference ``known'' sample and were not labelled as such in \Talk.
However, we did exclude these candidates when compiling the list of new
\sw lens candidates, as described in \Sref{sec:results:newcand}.


\subsection{Image presentation in \sw}
\label{sec:data:impres}

In order to perform a blind lens search over the entire \cfhtls
WIDE, we present the volunteers with cutouts of images selected
from the survey region. We briefly describe the image presentation here
for completeness; more information can be found in \PaperOne. We use
the $g$, $r$ and $i$-band imaging from \cfhtls which are most useful for
visual identification of lenses.  We made colour composite images using
the publicly-available code, \humvi\footnote{The open source  colour image
composition code used in this work is available from
\texttt{http://github.com/drphilmarshall/HumVI}} following the
prescription of \citet{Lupton2004}. The colour scales were chosen to
maximize the contrast between faint extended objects and bright early
type galaxies. These parameters
were then fixed during the production of all the tiles, in order to allow
straightforward comparison between one image and another, and for
intuition to be built up about the appearance of stars and galaxies
across the survey.

We extracted contiguous cutouts of size 82\arcsec\ (440 pixels), including
overlapping region of 10\arcsec (54 pixels) between the neighbouring
cutouts. This resulted in a catalog of some 430,000 cutouts for the entire \cfhtls
WIDE region. The size of the individual cutout was determined by
optimising factors such as the typical angular scales of gravitational
lenses, the number of objects seen in a single cutout and the total
number of image cutouts in the survey. If a lens
candidate happens to be too close to the edge of a cutout, then the overlap
between neighbouring cutouts allows a volunteer to get a clearer view
of the same candidate in at least one of the cutouts. We note that since
the images are shown randomly, a volunteer may not necessarily come
across the neighbouring cutout unless they classify a large
number of images. This is not a problem since our user base is extremely
large and we receive multiple classifications for the same cutout.


\section{Training sample}
\label{sec:ts}

The simulated lenses are important to train citizen scientists who may be
new to the task of finding lenses, but they are also crucial for
analysing the classifications performed by the citizen scientists (more
details can be found in \PaperOne, but see \Sref{sec:swap} below for a brief
summary). In this section, we describe the framework used for generating the
simulated lens sample, give details of the sample itself along with
some of its known limitations and also describe the sample of duds and
impostors.

\begin{figure*}
\begin{center}
\includegraphics[scale=1.0]{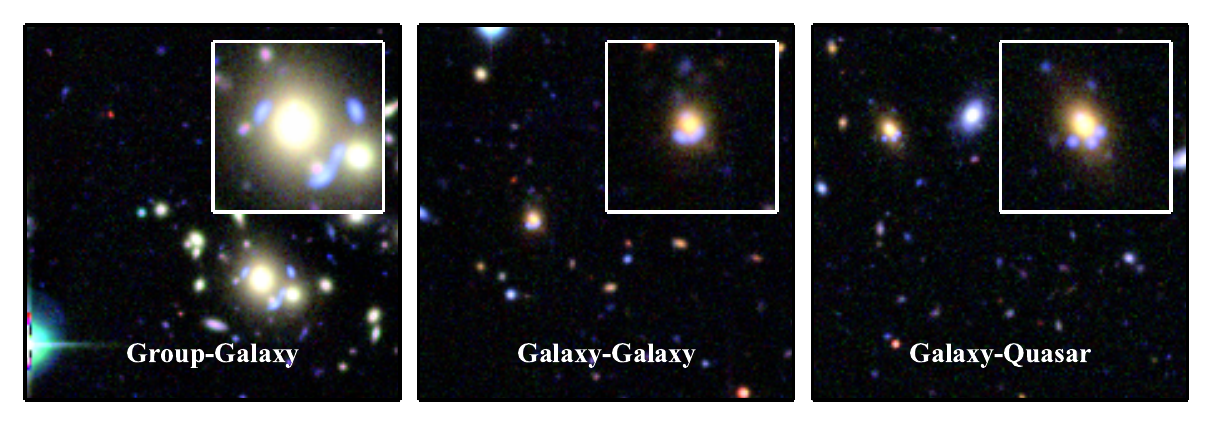}
\caption{ \label{fig:sim}
Examples of the three types of simulated lenses.
}
\end{center}
\end{figure*}

\subsection{Methodology to simulate lenses}
\label{sec:simmethod}

For the purpose of generating simulated lens systems, we divide them
into two main categories a) galaxy scale lenses b) and group or
cluster-scale lenses. We further subdivide galaxy scale lenses based on the
type of the background sources, namely galaxies and quasars. We do not simulate
group-scale  quasar lenses as they are expected to be even more rare. We now
describe our procedure to generate these different types of lens systems.

\subsubsection{Galaxy-scale lenses}
\label{sect:gallens}

We begin by considering all elliptical galaxies at $z<1$ in our parent
\cfhtls catalog \citep[]{Gavazzi2014} as potential lens candidates for
the simulated sample. To avoid using a known lens galaxy for our
simulation purpose, we exclude all those galaxies whose positions match
with the lensing galaxies from the SARCS samples within
2~arcsec\footnote{ Due to inaccuracies and uncertainties in measurements
of the centres of some of the lens candidates, some simulated lensed
images were superposed on the galaxies of known lens candidates. This issue was overcome by
presenting the same \cfhtls images with and without the simulated
lenses to the volunteers.}.

For each galaxy, the average number of source objects (either quasars or
galaxies) above a minimum luminosity $L_{\rm min}$ in the background that may get lensed
can be calculated as
\be
\label{eqn:nsrc}
N_{\rm src} = \int_{z_l}^\infty n_{\rm src}(>L_{\rm min},z_s)  \sigma_{\rm
lens}(\sigma_v,z_l,z_s,q) \frac{\rm d V}{{\rm d}z_s}
{\rm d}z_s
\ee
where
\be
\label{eqn:nlum}
n_{\rm src}(>L_{\rm min},z_s)= \int_{L_{\rm min}}^\infty \Phi(L',z_s) {\rm d}L' \,.
\ee

Here, $\Phi(L',z_s)$ denotes the source luminosity function per unit comoving
volume, $\sigma_{\rm lens}$ denotes the angular lens cross-section, which depends upon
the lens redshift ($z_l$), source redshift ($z_s$), the lens velocity
dispersion $\sigma_v$ as well as the projected axis ratio of the lens
ellipticity, $q$.

In order to calculate the lensing cross-section, we first calculate the
luminosity of each potential lensing galaxy using the photometric redshifts
($z_l$) from the parent galaxy catalog. Next, we use the $L-\sigma$ scaling relation from the
bright sample of \citep{Parker2005} given by
\be
\label{magstar2}
\sigma_v=142 \left(\frac{L}{L_{*}}\right)^{1/3} \, {\rm Km~s^{-1}}.
\ee
This sets the velocity dispersion of the halo
hosting the galaxy, which will be later used in the model.
We assume that the knee of the luminosity function of galaxies, $L_*$ evolves
such that there is a decline of $1.5$ magnitudes between $z=1$ to $z=0$
\citep{Faber2007}.\footnote{We anchor our $L_*$ evolution at low redshifts using the
determination of $L_*$ in the r-band by \citet{Blanton2001}. To maintain
consistency in magnitude systems, we have converted the CFHT MegaCAM magnitudes
to SDSS magnitudes and k-corrected them to $z=0.1$.}

We adopt a singular isothermal ellipsoid (SIE) model for each of our galaxies
\citep{Kormann1994}, such that the convergence is given by
\be
\kappa (x,y) = \frac{b \sqrt{q}}{2}\frac{1}{\left( \theta_1^2 + q^2\theta_2^2\right)^2} \,.
\ee
Here, $b$ is called the Einstein radius, and its dependence on the velocity
dispersion of the SIE is given by
\be
b = 4\pi\,
\left(\frac{\sigma_v^2}{c^2}\right)\left(\frac{D_{ls}}{D_{s}}\right) \,.
\ee
The SIE model results in a caustic and a pseudo-caustic on the source
plane: which demarcate the regions of different image multiplicities. We
make use of the parametric solutions, $r(\theta)$, for the caustics in
such a model from \citet{Keeton2000b} where $\theta$ is the polar angle.
We take the maximum of the radial and tangential caustic at every polar
angle in order to obtain the area of the lensing cross-section,
$\sigma_{\rm lens}$, for every galaxy,
\be
\sigma_{\rm lens}=\frac{b^2 q}{2} \, \int_0^{2\pi} r^2(\theta) d\theta \,.
\ee

We also add external shear at the centre of the potential lensing galaxy
drawn randomly from a set range (see \Tref{tab:thresh}. The shear is
expected to affect the lens cross-section for a small number of cases
when the shear strength is high in addition to high lens ellipticity or
the PA of the shear is almost orthogonal to that of the lens
ellipticity. However, the effect of shear on the lens cross-section is
expected to be small for most of the cases and is ignored in the current
implementation of SIMCT.

The luminosity functions of the background galaxies and quasars are
determined as follows. We use the results of \citet{Faure2009} to specify the luminosity
function of galaxies where the redshift distribution of sources is
given by
\be
\label{eqn:ps}
p_s=\frac{\beta z_s^2 {\rm exp}({\frac{z_s}{z_0(m_{\rm lim})}})^\beta}{\Gamma(3/\beta)z_0^3(m_{\rm lim})}
\ee
where $\beta=3/2$ and $z_0(m_{\rm lim})=0.13~m_{\rm lim} - 2.2$ and the source
counts as a function of the limiting magnitude are given by
\be
\label{eqn:ns}
n_s=\int^{m_{\rm lim}}_{-\infty} \frac{n_0 {\rm d}m}{\sqrt{10^{2a(m_1-m)}+10^{2b(m_1-m)}}}
\,,
\ee
with parameters $a=0.30$, $b=0.56$, $m_1=20$ and $n_0=3\times10^3~deg^{-2}$.

For quasars, we assume the luminosity function prescription of \citet{Oguri2010}
and adopt k-corrections by \citet{Richards2006}.

The luminosity function is expressed as
\be
\frac{{\rm d}\Phi}{{\rm d}M}=\frac{\Phi_{*}}{10^{0.4(\alpha+1)(M_{\rm abs}-M_{*})} + 10^{0.4(\beta+1)(M_{\rm abs}-M_{*})} }
\ee
where the normalization, $\phi_{*}=5.34\times10^{-6} h^3$ Mpc$^{-3}$ and break
magnitude, $M_*=-20.90 + 5 {\rm log} h - 2.5 {\rm log} f(z)$. The redshift
dependent factor in $M_*$ is given by
\be
f(z)=\frac{e^{\zeta z_s}(1+e^{\xi z_*})}{(\sqrt{e^{\xi z_s}}+\sqrt{e^{\xi z_*}})^2} \,.
\ee
We adopt the best-fit values $\zeta=2.98$, $\xi=4.05$, $z_{*}=1.60$
\citep{Oguri2010}. For the faint end slope, we use $\beta=-1.45$ whereas for
the bright end slope, we use $\alpha=-3.31$ when $z_s<3$ and $\alpha=-2.58$ at
higher redshifts, as prescribed by \citet{Oguri2010}.

With the cross-section, and the luminosity functions specified, we
calculate the expected number of sources behind a candidate lensing
galaxy using \Eref{eqn:nsrc}. We need to generate a large number of
simulated lenses (larger than the number of real galaxy lenses we expect
to find in \cfhtls) in order to have a reasonably large and diverse
training sample for thousands of \sw volunteers. Therefore, we
artificially boost the average number of sources by a factor (see
\Tref{tab:thresh}), which increases the occurrence of lensing. We
draw a Poisson deviate, $N_{\rm src}$ with a mean equal to the boosted
average number of sources. If $N_{\rm src}$ is greater than zero, then
this galaxy is flagged as a potential lensing galaxy.

Next, we determine properties of the background sources for every lens system.
We follow similar procedures for both background galaxies and quasars.
We draw source redshifts and luminosities from the aforementioned
distributions. We note that the sources are being drawn from a much
fainter magnitude range compared to the limiting magnitude of the
\cfhtls imaging and thus, the magnification bias\footnote{ In a
flux-limited sample from a survey, sources fainter than the flux limit
end up in the sample owing to the magnification by lensing which is
known as the magnification bias. This affects the source luminosity
function and needs to be accounted for when comparing the true and
observed luminosity functions.} is naturally taken into account. The
source positions with respect to the lens are drawn randomly from an
area inside the caustic.  When populating the sources within the
caustics, the finite size of the background galaxies is expected to
affect the lens cross-sections to some extent. As this factor is not
critical for the purpose of our training sample, for simplicity we assume the
background galaxies to be point like when computing cross-sections. We perform
ray-tracing for all of the $N_{\rm src}$ sources using the publicly
available code \gravlens \citep{Keeton2000} and choose sources that
satisfy our selection criteria given below. We determine the fluxes of the
lensed images and the total magnification of each of the lensed sources.
We draw a source randomly for which the flux of the second brightest
lensed image and the total magnification of all lensed images meet the
thresholds given in \Tref{tab:thresh}.

Since we want to produce realistic looking lens systems, we simulate
lenses in each of the five \cfhtls~filters. The colours of the background
galaxies are drawn randomly from the photometric CFHTLenS catalog
\citep{Hildebrandt2012,Erben2013}.  Similarly, we use a quasar catalog
from the SDSS Data Release 9 \citep{Paris2012} from which colours are
drawn to simulate quasar lenses. Next, we assume a Gaussian
profile\footnote{This was due to an oversight. We intended to use either
an exponential or de Vaucouleurs' profile that will be adopted for
future implementations of SIMCT.} for the galaxies. The ellipticity and
the position angle (PA) are drawn
randomly from within the range given in \Tref{tab:thresh}. The
effective radius of the galaxy is estimated from the luminosity$-$size
relation \citep[][with a redshift scaling, to account for size evolution]{Bernardi2003} given by
\be
R_{\rm eff}= 10^{0.52} \frac{L_r^{2/3}}{{(1+z_s)}^2} \, {\rm Kpc}
\ee
where $L_r=L_s/10^{10.2} L_\odot$. On the other hand, quasars are assumed to be
point sources and the PSF, with which quasars are convolved, is assumed
to have a Gaussian profile. The full-width-at-half-maximum of the PSF is
equated to that of the mean seeing for every filter. The mean seeing
values are taken from Table 4 of the official Terapix T0007 release
explanatory document \footnote{
\texttt{http://terapix.iap.fr/cplt/T0007/doc/T0007-doc.pdf}}.

Once all the parameters are determined for the lens and source models,
we once again use \gravlens to generate simulated lensed images.  After
accounting for the shot noise in the lensed images and convolving them
with the seeing in each of the filters, the simulated image is added to
the real \cfhtls~image centred on the galaxy chosen to act as a
lens. Note that we ensure that the lensed galaxies and lensed quasars
are not superposed on the same ``lensing'' galaxy.  Similarly,
these ``lensing'' galaxies at the galaxy scales are ensured to be
distinct from those chosen for the group scales. The framework for the
group-scale is described below.

\begin{figure}
\begin{center}
\includegraphics[scale=1.2]{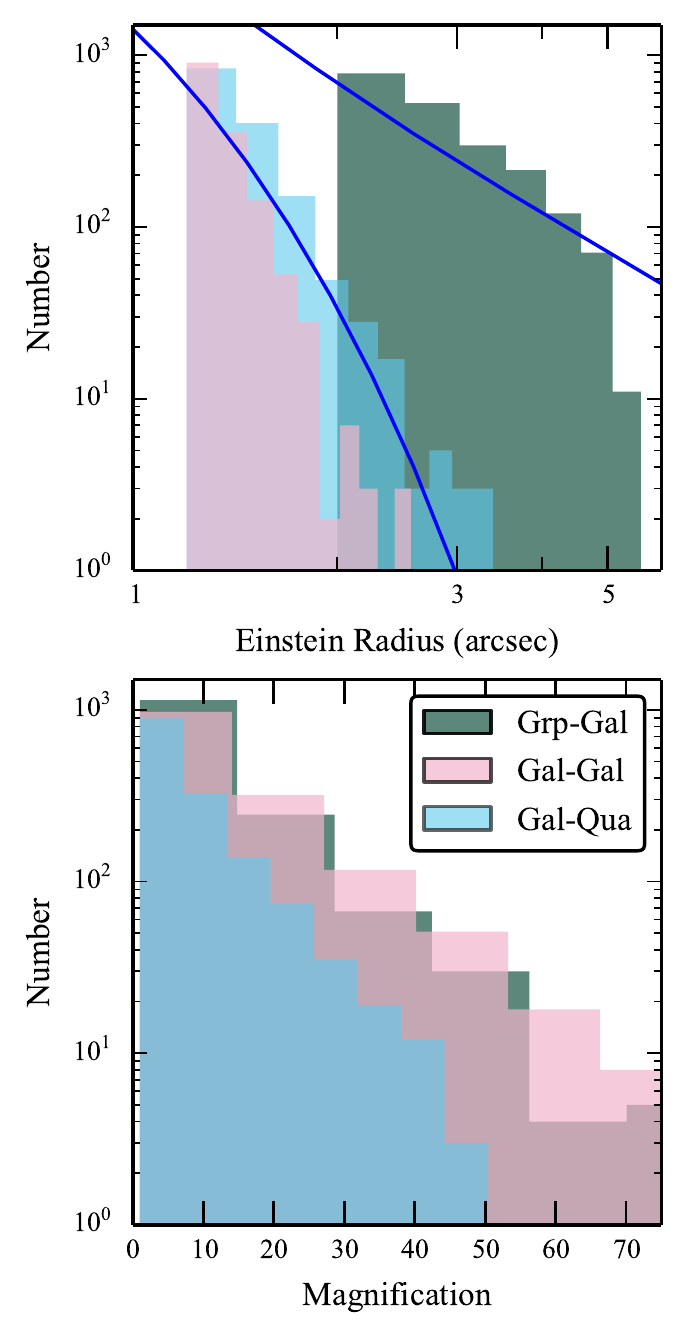}
\caption{ \label{fig:remudist}
Einstein radius and total magnification distributions for all types of lenses. The solid (blue)
curves show the theoretical prediction assuming an SIS model at galaxy-scales
and a total (NFW+Hernquist) model at group scales taken from \citep{More2012}.
}
\end{center}
\end{figure}

\begin{table}
\begin{center}
\caption{ \label{tab:thresh}
Thresholds used in the selection of the simulated lenses. }
\begin{tabular}{l l l l l}
\hline
Name  &  \multicolumn{2}{c} {Gal-Gal (Grp-Gal)}  & \multicolumn{2}{c}{Gal-Qua} \\
      & min  &  max  & min & max \\
\hline
\hline
Source Redshift  & 1.0 & 4.0  & 1.0  & 5.9 \\
Source Flux      & 21.0 & 25.5 & 21.0 & 25.5 \\
Source ellipticity & 0.1 & 0.6 & $-$ & $-$ \\
Source PA & 0 & 180 & $-$ & $-$ \\
Lens Redshift  & $-$ & 0.9  & $-$  & 0.9 \\
Lens shear strength &  0.001 ($-$) & 0.02 ($-$) &  0.001 & 0.02 \\
Lens shear PA &  0 ($-$) & 180 ($-$) & 0 & 180  \\
Einstein radius (arcsec) & 1.2 (2) & 5 ($-$) & 1.2 & 5 \\
\hline
boost factor     & $=$100 (40)  &  & $=$1200 & \\
Image Flux$_{\rm 2B}$ & $>$23  & & $>$23 & \\
Image Flux$_{\rm tot}$ & $<$19 & & $<$20 & \\
\hline
\end{tabular}
\end{center}
a) () -- corresponds to quantities used for Grp-Gal scale lenses, if they
are different from Gal-Gal.
b) $_{\rm 2B}$ --  the second brightest lensed image.
c) $_{\rm tot}$ -- total flux integrated over all of the lensed images.
d) All fluxes are in AB mag. PA is in degrees measured East of North.
\end{table}

\subsubsection{Group-scale lenses}

At group or cluster-scales, the mass distribution is more complex. The
convergence in the inner regions, which are typically responsible for the
multiple lensed images, arises from not only the brightest group galaxy (BGG) at
the centre, but also from the dark matter component and
the satellite galaxies \citep{Oguri2005,Oguri2006}. We generate a basic group
catalog based on the magnitudes and photometric redshifts available for the
\cfhtls. We select all galaxies with 10$^{10.8} M_\odot$ as plausible BGGs. We
select the member galaxies such that their photometric redshifts are within
$\delta z = 0.01$ of the BGG and within an aperture of $250$~Kpc. If another BGG
is found within the aperture, then the fainter BGG is removed from our list of
BGGs.

\begin{figure*}
\begin{center}
\includegraphics[scale=1.3]{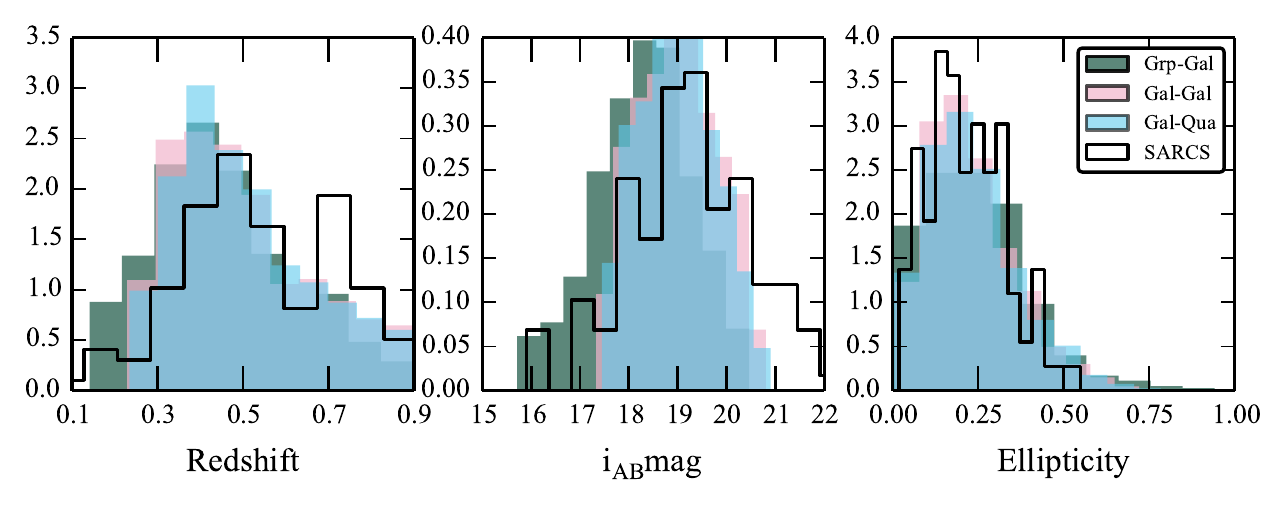}
\caption{ \label{fig:lensprop}
Distributions of properties of the ``lensing'' galaxies of the simulated
sample compared to the known lens sample SARCS.
}
\end{center}
\end{figure*}

We assume a constant mass$-$to$-$light ratio of $3 \times
0.7~h~ M_{*}/L_{*}$, to convert the BGG luminosity to a stellar mass
estimate. The stellar mass$-$halo mass relation \citep{Behroozi2013},
including random scatter, is then used to calculate the halo mass for
the lens. We adopt an NFW \citep{Navarro1997} density profile for the
underlying dark matter halo. Given the halo mass, other key parameters
such as the scale radius ($r_s$) and the density at the scale radius
($\rho_s$) can be determined for an NFW profile. In addition, we adopt
an SIE model for the BGG and members whenever the ellipticities are
available from the galaxy catalog (else we use an isothermal sphere,
SIS).

We calculate the luminosities and velocity dispersions for the BGG and each
of the member galaxies following the same prescription as in
\Sref{sect:gallens}. To calculate the average number of sources that get
lensed by such a system, we need to calculate the lensing cross-section
for each of these potential lensing groups. The complexity in the lens
models makes it analytically intractable to calculate the size of the
caustics\footnote{The lens mass distribution determines size and shape
of the caustics. Any source located within the caustics will form
multiple lensed images which is the criteria for strong lensing. To
further understand caustics, see e.g., \citet{Schneider1992}.}.
Therefore, we generate the caustics numerically using \gravlens and then
determine the area covered by the caustics. We consider only galaxies as
our background source population since group or cluster-scale quasar
lenses are expected to be extremely rare in the \cfhtls.  Following the
same procedure as described in \Sref{sect:gallens}, we calculate the
number of galaxies expected to lie behind every potential lensing group
(see \Eref{eqn:nsrc}).  As before, for each background galaxy within the
lens cross-section, a redshift and an $i$-band magnitude is determined
by drawing galaxies randomly from the respective distributions (see
Equations~\ref{eqn:ps}-\ref{eqn:ns}).

All those groups that are found to have no background galaxies within the
cross-sectional area are rejected and the rest are included as potential lenses.
As mentioned earlier, we artificially boost the total number of sources behind
every lens but ensure that the statistical properties such as the profile of the
image separation distribution are not affected (see \Fref{fig:remudist}). We
follow the same procedure and apply thresholds to determine properties
of the lensed galaxies for every lens as described for galaxy-galaxy lenses
in the previous section. The thresholds are same as those used for
galaxy lenses (see \Tref{tab:thresh}) and are reported within ``()'', if
different for group scales. The simulated lensed images are then added
to the real \cfhtls~images with the BGGs as the centre by following
exactly the same procedure as described in the previous section.


\subsection{Simulated lens sample and catalog description}

In this section, we describe some of the properties of our simulated sample for
each of the three types of lens samples. We have made an attempt to
generate as realistic a lens sample as possible within the requirements
of the \sw project. The statistical properties of the lens sample are
expected to be similar to real lens samples.

In \Fref{fig:remudist}, we show the Einstein radius ($R_{\rm E}$) distribution for the
galaxy-scale and group-scale simulated lenses. For comparison, we give
the expected distributions (blue solid curves) for an SIS like density
profile at galaxy-scales and an NFW+Hernquist profile at group scales.
The theoretical curves are taken from \citet{More2012} wherein the
models are explained in detail. We note that the model we adopt at
the group scale also includes SIS or SIE components for the group members
unlike the theoretical prediction. The theoretical curves have arbitrary
normalizations. We also show the distribution of the total magnification
for all three samples.

Next, we consider the redshift, magnitude and ellipticity distributions
of the ``lensing'' galaxies from the simulated sample as shown in \Fref{fig:lensprop}. For reference,
we also show SARCS lenses from \citet{More2012}, with arbitrary
normalizations. We find that the properties of the foreground lenses in
the simulated and the real lens samples are broadly similar.

We produce catalogs with lens and source properties for each of the
three types of simulated lenses.  The catalogs typically have lens position,
redshift, magnitudes, Einstein radius, ellipticity (whenever available)
and shear (for galaxy-scale lenses only). For the background sources, we
provide the offset from the lens centre, redshift, magnitudes, total
magnification, number of lensed images. Additionally, when possible,
ellipticity and effective radius of the background galaxies have also
been provided. These catalogs are available from
\texttt{https://github.com/anupreeta27/SIMCT} and the
simulated lens image sample is available from the authors on request.


\subsection{Limitations of the simulated lens sample}

The simulated lens sample, although realistic, is not perfect, due to
the simplicity of the lensing models and our limited understanding of
the uncertainties in the model parameters. Comments from citizen
scientists were very helpful in order to identify some of these failures, which
make up roughly 5\% of the simulated sample.\footnote{This estimate is
based on the number of \#simfail tags from TALK, the discussion forum.}
Here, we describe some of the cases or aspects in which the simulations
were known to have failed to look realistic.

The parameters required by the models' various scaling relations
primarily depend on the photometry of the galaxies, groups and quasars
detected in the survey. For galaxy-scale lenses, the fainter or higher redshift
galaxies, chosen to act as lenses, tend to have poor photometric redshift
measurements. Consequently these galaxies were occasionally assigned the wrong
luminosity and velocity dispersion estimates, resulting in
simulated lenses which look implausible or unrealistic.  For example,
the lensed images for some of the failed simulations have larger image
separation than what one would expect from the luminosity and/or size of the
galaxy. We roughly expect mass to follow light, so more massive galaxies
typically look brighter and/or bigger.

At group scales, the magnitudes and photometric redshifts were used when defining
the group membership. Therefore, errors in redshift estimates occasionally
generated galaxy
groups having member galaxies with unrealistically dissimilar properties.
In some cases,
low redshift spiral galaxies were incorrectly assigned high redshift.
Spiral galaxies are typically less massive and low redshift spiral galaxies are
unlikely to act as gravitational lenses. Hence, some such instances
did not appear convincing, as the lensed images again did not have the expected
configurations or separations.

We also use a single component to describe the light distribution of the
background galaxies. This is clearly not the most accurate description
for galaxies, especially for the irregular star-forming galaxies which
comprise a significant fraction of the lensed galaxy population.
Star-forming galaxies have complex structures such as star forming
knots, spiral arms, bars and disks. The simulated lensed images do not
display these features. This is not problematic for most of the images
taken from ground based telescopes such as the CFHT but sometimes the
profiles of the lensed arcs can appear very symmetric (along the length
or width of the arc) and featureless, especially, if the images are very
bright.


\subsection{Duds and Impostors}
\label{sec:dfp}

Citizen scientists need training not only to identify gravitational
lenses, but also to reject images which either contain no lenses, or contain
objects which could be mistaken for lenses. Hence, in addition to the simulated
lenses, we added a sample of ``duds'' and ``impostors'' to the training
sample. Duds are images which have been visually inspected by experts
and confirmed to contain no lenses. Impostors are systems
which have lens like features but are not lenses in reality, for example, spiral galaxies,
star-forming galaxies, chance alignments of features arranged in a
lensing configuration and stars.

We selected a sample of 450 duds for the \StageOne classification in \sw
and a sample of 500 impostors for the \StageTwo inspection. The
sample of impostors was selected from the candidates which passed
the \StageOne of \sw. We note that this is the first time we have a
systematically compiled sample of visually inspected impostors by
the \sw volunteers and categorized by the science team. We produced an
additional larger sample of a few thousand false positive detections
by scanning through the
low probability images after the completion of \StageTwo. All of these
data products will be made available at
\texttt{http://spacewarps.org/\#/projects/CFHTLS/}. Such a sample has tremendous
utility for training and testing of various lens finding algorithms
\citep[e.g.,][]{Chan2014}.

\begin{figure}
\begin{center}
\includegraphics[scale=0.6]{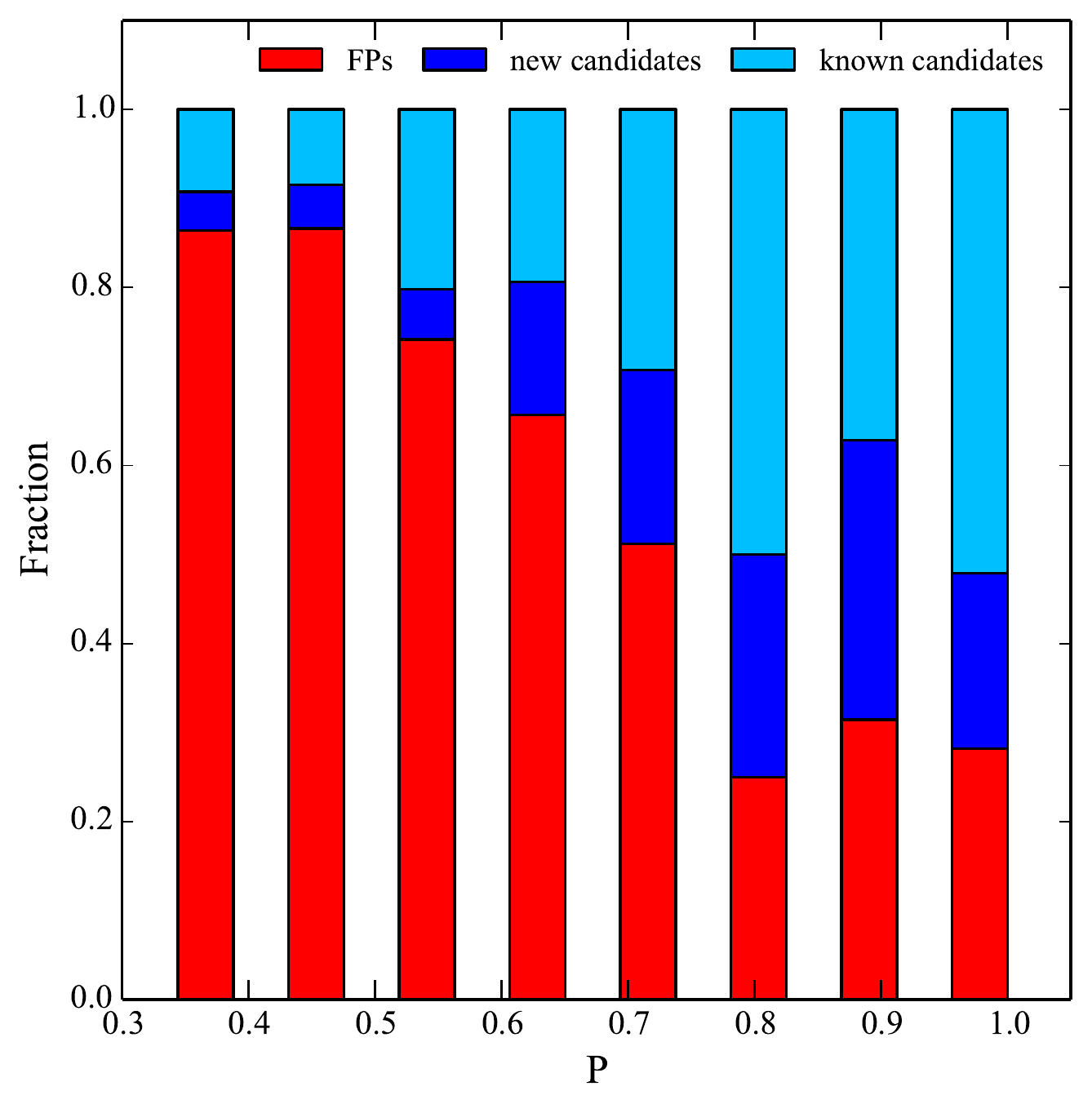}
\caption{ \label{fig:stackP}
Distribution of different types of candidates as a function of the
posterior probability $P$, obtained at the end of \StageTwo. The types of the candidates are the false
positives (FPs), the new candidates and the known candidates. The new
and known candidates have higher detection rate for higher values of $P$,
as expected.}
\end{center}
\end{figure}


\section{Methodology to produce the \sw-\cfhtls lens sample}
\label{sec:swap}

\sw works as a single unified system which uses the method of visual
inspection to find gravitational lenses. For the first \sw lens search,
the volunteers were shown images at two stages. At \StageOne, volunteers
were asked to carry out a rapid inspection to select lens candiates
ranging from possible lenses to almost certain lenses. At \StageTwo,
volunteers were asked to inspect the candidates from \StageOne and select
only promising lens candidates. A daily snapshot of the classifications
performed by volunteers was provided to the science team every night.
This daily batch was analysed by the Space Warps Analysis Pipeline
(SWAP). The philosophy and the details of SWAP are described in detail
in \PaperOne. Here, we briefly summarize how it works.

Each subject (image cutout) is assigned a prior probability of
$2\times10^{-4}$ of containing a lens system. Every volunteer
is assigned an agent characterised by a $2\times2$ confusion matrix~$\CM$,
which quantifies the volunteer's ability to correctly classify an image
as containing a lens ($\PL = P_L$) or not containing a lens ($\PD = P_D$).
The values of
these confusion matrix elements are determined based on the performance of the
volunteer on the training sample, specifically, $P_L$ (and $P_D$) is
determined based on the fraction of simulated lenses (and, respectively, duds)
correctly classified. After every classification, the agent updates the
probability of the classified subject based on the volunteer's classification
and the confusion matrix, according to Bayes' theorem.
The agent's confusion matrix is updated after the
classification of every training image. The thresholds for the
probabilities to accept or reject a subject if it contains a lens or does
not contain one can be chosen in SWAP. In \StageOne, those images which cross
these threshold values are ``retired'', and are not subsequently shown to the
volunteers. In this way, the crowd can use its time efficiently in
inspecting previously unclassified subjects.

SWAP was run nightly during \StageOne in order to retire subjects and inject new
ones in to the classification stream. The subjects that passed the detection
threshold at the end of \StageOne were served again at \StageTwo for careful
re-inspection. The goal of \StageOne inspection was to maximise completeness
and that of \StageTwo was to maximise purity. Each subject, after \StageTwo, has a final posterior
probability $P$. In the ideal case, all images containing lenses will have high $P$
values and those without lenses will have low $P$ values. In practice, we expect a
small fraction of the real lenses (or non-lenses) to be assigned low (or high) $P$
values, thereby decreasing the completeness (or purity) of the final lens
sample. As this is the first lens search with \sw, we want to find a
threshold $P$ value which will result in acceptable levels of completeness and
purity of the final sample of lens candidates.

To achieve this, we selected a total of 665 subjects with $P>0.3$
at \StageTwo which were then visually inspected by three of us (as ``lens experts'',
 AM, AV, PJM).  Each images was assigned a grade on a scale of 0 to 3,
representing those images
0) unlikely to contain a lens, 1) possibly containing a lens, 2) probably
containing a lens and 3) almost certainly containing a
lens. The final sample of \sw-\cfhtls lens candidates was
then produced by selecting candidates above a threshold on the
averaged grade~$G$, as described in the next section.

\section{Results}
\label{sec:results}

\subsection{\sw-\cfhtls lens sample}
\label{sec:swlens}

\begin{table}
\begin{center}
\caption{ \label{tab:stats}
Statistics of detections in \sw }
\begin{tabular}{l l l l l}
\hline
   &  \multicolumn{2}{c} {\StageOne}  & \multicolumn{2}{c}{\StageTwo} \\
      & KC  &  KL  & KC & KL \\
\hline
\hline
Number  & 142 & 39 & 79  & 34  \\
Fraction  & & & & \\
of total recovered & 32\% & 66\% & 18\% & 58\% \\
$P_{\rm acc thresh}$ & 0.95 & 0.95 & 0.3 & 0.3 \\
Averaged Grade (G) & $-$ & $-$ & 1.3 & 1.3 \\
\hline
   &  \multicolumn{2}{c} {\StageTwo}  &   & \\
      & NC  &  AC  &  & \\
\hline
\hline
Number  & 59 & 141 &  & \\
Fraction & & & & \\
of detections & 14\% & 33\% &  & \\
$P_{\rm acc thresh}$ & 0.3 & 0.3 &  & \\
Averaged Grade (G) & 1.3 & 1.3 & &  \\

\hline
\end{tabular}
\end{center}
{KC}-- Known lens candidates \\
{KL}-- Known lenses \\
{NC}-- New lens candidates  \\
{AC}-- All (known and new) lens candidates  \\
$P_{\rm acc thresh}$ -- systems with Posterior probability~$P$ above this threshold are selected \\
Note: For KC and KL, Percentages are with respect to the known
population (i.e. 438 KC and 59 KL see \Sref{sec:data:kls}) whereas for
NC and AC, percentages are with respect to the total sample of 429 lens
candidates.
\\
\end{table}

In this section, we describe the \sw candidate lens sample from the
\cfhtls.  We find a total of 141 candidates with $G\ge1.3$ (medium-high
grade), of which 59 are new systems. This sample is further divided as
follows. We have a total of 50 candidates with $1.3\le$G$<2$ (medium
grade), of which 30 are new. The quality of candidates in this category
is such that at least one of the inspectors (``lens expert'') thought the candidate was
probably a lens (that is, a grade of 2) and a second inspector thought
that it was possibly a lens (that is, a grade of 1). Among our high
grade sample ($G\ge$2), there are a total of 91 candidates, of which 29
are new. In this category, the minimum grade by all of the inspectors
was 2, suggesting that the candidates are probably or almost certainly
lenses according to all three inspectors. To avoid duplication, only the
newly discovered lens candidates with G$\ge 1.3$ (medium-high
probability) are presented in this paper (see
\Sref{sec:results:newcand}), and further information on \sw-selected
candidates that were previously identified in the literature (as
described in \Sref{sec:data:kls}) will be made available at
\texttt{http://spacewarps.org/\#/projects/CFHTLS/}.

We also find a total of 288 (and 245 new) candidates with averaged grade
$0<$G$<1.3$ (low grade), which means that at
least one of the inspector thought the candidate was possibly a lens, and
in the best cases, all three inspectors thought the candidate was possibly
a lens (that is, a grade of 1)\footnote{If grades from the inspectors
were found to be discrepant by 2 or more, these were discussed and
re-graded to resolve the discrepancy.}. Further information on the low
probability sample such as their positions and images will be available
at \texttt{http://spacewarps.org/\#/projects/CFHTLS/}. Note that if all of the
inspectors gave a grade of 0 to a candidate, then it was discarded from
the sample.

In \Tref{tab:stats}, we give overall statistics of the systems detected
at \StageOne and \StageTwo. We give the total number of detections of
the known lens candidates, known lenses and the new lens candidates at
each stage.  We also give the recovery fractions for the known samples
and fraction of total detection for the new samples at each stage.
Overall, the sample of new \sw lens candidates comprises over
40~per~cent of the total \sw$-$\cfhtls lens candidates. We find that
90~per~cent of the confirmed lenses found at \StageOne are also
recovered at \StageTwo.  However, $\sim35$~per~cent of the known lenses
were missed already at \StageOne: we return to the discussion of these
false negatives below in \Sref{sec:fn}. Nearly 45~per~cent of the known
lens candidates from \StageOne are rejected at \StageTwo. Such fractions
are acceptable for ``candidates'' as their quality grades vary from high
to low.  

In \Fref{fig:stackP}, we plot the distribution of false positives and
the high grade lens candidates, as a function of the $P$ value assigned
by SWAP at the end of \StageTwo.  On average, the fraction of lens
candidates is indeed an increasing function of $P$. This shows that the
\sw generated $P$ values for the subject are roughly correlated with the
expert grades albeit with quite some scatter.  We note that below
$P\sim0.75$, the fraction of false positives starts to exceed the
fraction of real lens candidates. This could be a good threshold to
choose to maximize the purity of the final sample. However,
choosing $P=0.3$ gives a completeness of 92\% for the ``known lens''
sample instead of 64\% for $P=0.75$. Therefore, the new sample should
also have increased completeness; expert grading then allows us to
increase the purity of the sample.


\subsection{New lens candidates from \sw}
\label{sec:results:newcand}

We give basic information about the final sample of 59 new medium-high grade
lens candidates found by \sw in
\Tref{tab:swcands}. We report the candidates with a \sw ID and Name
of the lens system. We give their positions (RA, Dec), photometric
redshift ($z_{\rm phot}$), $i$-band magnitude of the lensing galaxy,
averaged grade~$G$ from the lens experts, zoo ID (identifier used in
TALK), $P$ value at \StageTwo and a visual categorization of the type of
lensed images and the lensing galaxy in the ``Comments'' column in
\Tref{tab:swcands}. Whenever available the lens properties are
taken from the \cfhtls photometric catalog \citep{Coupon2009}; otherwise,
for the lens galaxy positions, the reported values were measured
manually. The visual categorization of the lens type is only suggestive
and the explanation of the notations in the Comments column is given at
the bottom of the table.

\begin{figure*}
\begin{center}
\includegraphics[scale=1.9]{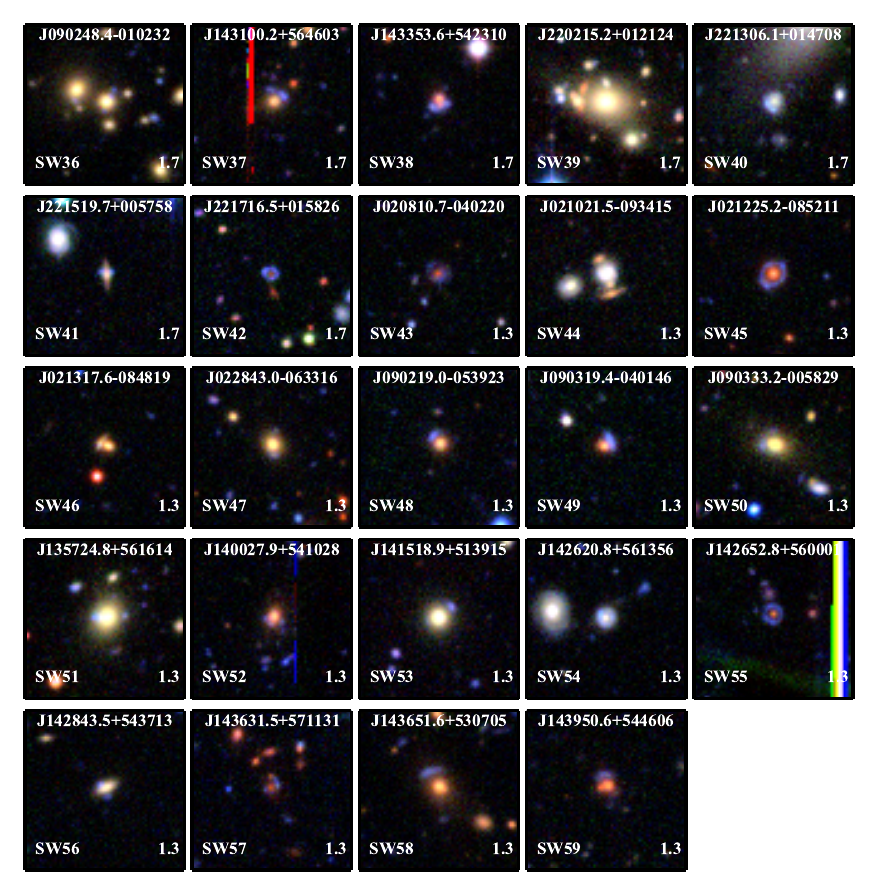}
\caption{ \label{fig:lc}
The new \sw lens candidates with expert grade G$>=$1.3. The images are
30\arcsec on the side.
}
\end{center}
\end{figure*}

We show images of our new sample in \Fref{fig:lc}. The panels are arranged
first in the descending order of their grades, and within each grade, in
ascending order of RA. As the first lens search was a blind search with no
pre-selection of candidates from any algorithm, we find various types of
lenses, as expected from such a search. The final sample consists of
both galaxy and group-scale lens candidates. There are detections of
elongated arcs and some interesting point-like quasar lensed images.
Most of them are brighter in the bluer $g$ band, but some candidates brighter
in the redder $i$ band are also found. Since the robotic lens searches
focused on the blue lensed features, they are likely to miss
such interesting lens candidates. We did not find any examples of
exotic lens candidates from the visually inspected $P>0.3$ sample.
There may be some more interesting candidates that were missed either at
\StageOne or \StageTwo but have been identified by volunteers in \Talk.
This resource is yet to be mined and is left for future work.

The new \sw lens sample presented here illustrates some of
the advantages of having
citizen scientists find lenses through visual inspection. An algorithm,
by definition will find objects that adhere to a selection criteria that
uses either geometry or flux information from an image. On the other
hand, citizen scientists can interpolate over or
extrapolate beyond the basic selection criteria provided to them.  For
example, the lower blue arc in SW7 is split by a small red galaxy.  An
algorithm typically fails to detect such arcs because the arc is broken
into smaller arclets which then falls below the minimum length or area
allowed for an arc to be detected. Human inspectors have no
problem in interpolating over the broken blue arc over the red galaxy,
understanding that it is a single long arc. The system SW20 has
point-like lensed images which cannot be detected by arc finding
algorithm, whereas the ring finding algorithm may have missed this because of
the atypical colour and structure of the lensing galaxy. Detection of red arcs, for
example, as seen in the SW39 candidate, shows how the volunteers
extrapolate on the colour parameter: the training sample contains
predominantly blue arcs, because the source colours were drawn from realistic
observed distributions.

The power of citizen scientists also lies in the high dynamic range that
allows us to find systems which have very short (thick) to long (thin)
arcs, from highly compact to low surface brightness images, from round
and point-like to elongated and curved images, from blue to red, from
regular to exotic kinds of lenses; while keeping the false positive rate
low compared to algorithms. Discovery of this large sample of completely
new candidates missed by some of these algorithms demonstrates that the
\sw system is functioning well, the self-taught citizen scientists
reaching parts of discovery space that the robots did not.

Further detailed qualitative and quantitative analysis of the properties
of the entire \sw sample (new and previously identified candidates) and
the mass modelling analyses for the new candidates will be presented in
a subsequent \sw paper (Verma et al., in prep.).

\begin{figure*}
\begin{center}
\includegraphics[scale=1.9]{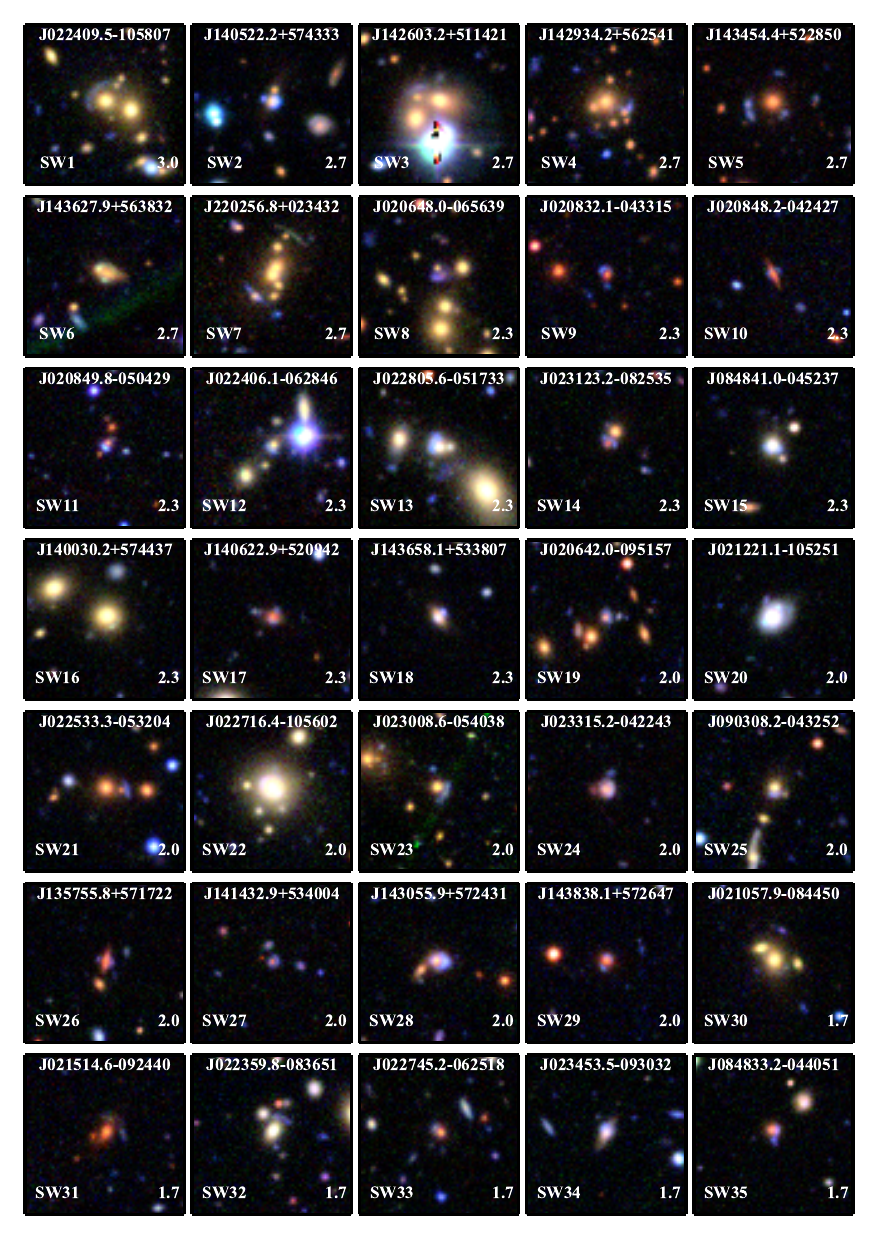}
\end{center}
\end{figure*}

\subsection{Measurements of properties of the lens and the lensed images}
\label{sec:results:meas}

In the subsequent sections, we compare various properties of the lens
candidates. Here, we describe how we extract or measure these properties, namely,
the lens redshift, the Einstein radii and the total flux of the lensed images or
arcs.

We use the publicly available redshifts for the lens galaxy from the
\cfhtls photometric catalogs \citep{Coupon2009}. The total flux of the
lensed image or arc is measured in the $g$-band but the adopted method
is different for different samples. For the simulated sample, we
multiply the magnification of the second brightest image with the source
magnitude. For the \rf sample, the arcs are detected in the scaled
difference image of $g$ and $i$-bands from which the lensing galaxy is
subtracted \citep[for details, see][]{Gavazzi2014}. Here, we use the flux
of the lensed images measured by \sextractor from the scaled
difference image, that is, $g-\alpha i$ and convert it to the $g$-band
flux using mean colours of the foreground and background population. For
the \af and the \sw sample, we integrate the flux in the image pixels
identified by \af or \sextractor.

\begin{figure*}
\begin{center}
\includegraphics[scale=1.0]{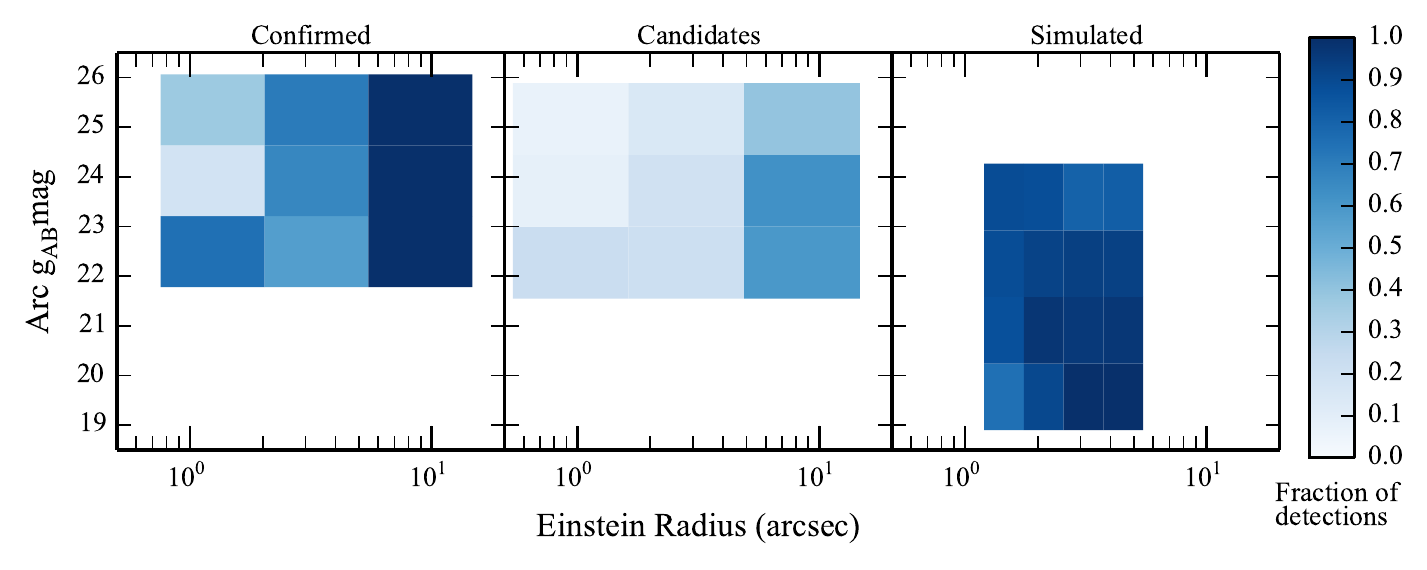}
\caption{ \label{fig:compre} Fraction of lens candidates recovered by \sw as a
function of the arc magnitude ($g$ band) and the Einstein radius for three lens
samples, namely, the known lenses, the known lens candidates and the simulated
sample. }
\end{center}
\end{figure*}

The Einstein radius is also measured differently for
different samples. For the galaxy-scale lenses in the simulated sample,
we use the value of the input parameter of the lens model for the $R_{\rm E}$.
For group-scale lenses, since the lens model is multi-component, we
need to determine the $R_{\rm E}$ from the image positions. We use those
pairs of lensed images that have the smallest and the largest angular
separations. The $R_{\rm E}$ here is then half of the averaged values of
these angular image separations. For the \rf sample, we use the peak
position of the lensed images measured by running \sextractor on
the scaled difference image. We calculate the image separation from the
lens centre as a rough estimate of the $R_{\rm E}$. For the \af (SARCS)
sample, we use the same definition as above except that the peak
position is identified either by the \af or manually. For the \sw lens
sample, the same definition is used where the peak positions are
identified either with \af or \sextractor.


\subsection{Recovery of known lens samples from the \cfhtls by \sw}
\label{sec:results:known}

We now determine the fraction of the known sample of lenses that were
recovered by \sw. Note that this sample corresponds to the \rf and \af
samples combined, as defined in \Sref{sec:data:kls}. In
\Tref{tab:stats}, we show that $\sim32\%$ of the known lens candidates,
and $\sim65\%$ of the known lenses were found at \StageOne. We find that
$56$\% of the known lens candidates and $87$\% of the known lenses from
\StageOne passed our \StageTwo selection criterion of $P>0.3$ and
$G>1.3$. The left and the middle panels of \Fref{fig:compre} show the
fraction of detections as a function of arc magnitude and the Einstein
radius of the lens systems for the known confirmed lenses and lens
candidates. As expected, we find that systems with brighter images
and/or with larger $R_{\rm E}$ are detected more often in \sw.

We find that most of the confirmed lenses and candidates that are missed
by \sw are systems from the \rf sample, with fainter arcs and smaller
$R_{\rm E}$.  The main reason why \rf found such candidates is because
it involves subtracting the light from the lens galaxy, making it easier
to detect the lensed images during both the automated object-finding,
and the visual inspection and classification phases. This approach
naturally improves the detection efficiency at smaller $R_{\rm E}$ and
for fainter systems. The \sw volunteers were not shown any
galaxy$-$subtracted images. Showing galaxy$-$subtracted images might be
a better strategy to adopt for future lens searches at galaxy-scales
with \sw. However, we note that accurate modelling and subtraction
of the galaxy light profile in different bands is challenging and better
techniques are being actively developed to enhance detections of lenses
at small image separations. In \Sref{sec:fn} below, we further explore
and discuss why some of the confirmed lenses were missed by \sw.


\subsection{Image separation distribution}
\label{sec:results:isd}

The distribution of image separations (i.e. 2~$R_{\rm E}$)
can be used to probe the average density profile of the lens population
\citep{Oguri2006,More2012}.  However, the lens sample found by the \af
may have incompleteness as a function of the image separation. Thus, the
lack of understanding of the selection function of the lens sample may
affect the constraints on the density profile. A blind lens search done
by visual inspection alone, for example, through \sw citizen scientists may find
lenses missed by the \af search and thereby, improve completeness.

\begin{figure}
\begin{center}
\includegraphics[scale=1.2]{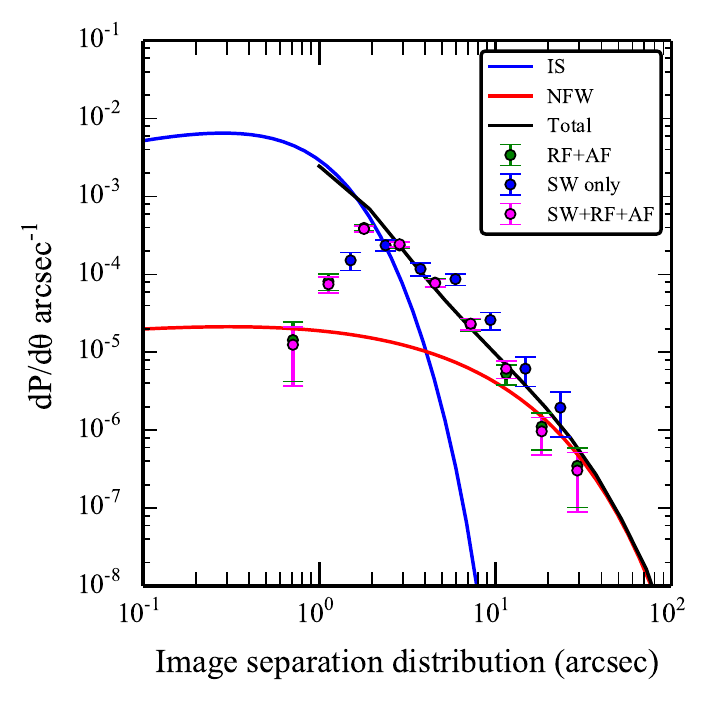}
\caption{ \label{fig:isd} Image separation distribution (ISD). Comparing theoretical
predictions (solid curves) with the \cfhtls known lens samples (green)
and the combined sample of known and \sw lens candidates (magenta). The
sample of new and the known lens candidates discovered from \sw alone is
shown in blue. The new updated profile of the ISD (magenta) is
consistent with our previous measurements and strengthens our conclusion
that the average density profiles of the lenses are similar to the Total
profile.}
\end{center}
\end{figure}

Indeed, we have found 59 new medium-high grade lens candidates that were
not known before. In \Fref{fig:isd}, we show the image separation
distribution using all the known and new lens candidates. The different
data points are  the known \rf and \af sample (green), the \sw
identified (known and new) lens sample (blue) only and the combined
\cfhtls sample of \rf, \af and the new \sw lens sample (magenta). It is
interesting to note that both the \rf$+$\af and \sw samples have very
similar profiles and thus, the profile of the combined sample has not
changed significantly. This implies that previous constraints on the
image separation distribution are robust and the \af selected sample
does not suffer from significant incompleteness for medium to large
$R_{\rm E}$. This is the regime that probes density profiles of galaxy
groups to clusters. 

In the figure, we also show for comparison the theoretical predictions
corresponding to three density profiles, namely, isothermal sphere (IS),
NFW and a ``Total'' profile which has NFW and Hernquist profiles
combined with an adiabatically contracting model for the dark matter
component \citep{Gnedin2004}. These curves are taken from
\citet{More2012}, which gives details of the calculation of these
predictions.  With the updated sample of lens candidates, we confirm our
previous prediction that the mass density profiles of galaxy groups is
indeed consistent with the ``Total'' profile.  At smaller image
separations ($\lesssim 2$~arcsec), the ``Total'' profile converges to
isothermal like case and assuming these predictions are reliable, we
find that the lens samples have very low completeness. This is not too
surprising compared to the 40\% completeness expected for the \rf sample
\citep{Gavazzi2014}.


\section{Discussion}
\label{sec:discuss}

Finding gravitational lenses is a difficult and complex task. No single
method is perfect, each method has some advantages over the other. It
may be the case that a single method may never be the best method for
optimising completeness and purity. Visual inspection will likely be
required for pruning candidates at some stage of lens candidate
selection even in the future. Therefore, we would like to understand how
best we should combine the strengths of robots and humans to optimize
the lens finding method.

In this section, we first compare the lens candidates found by
\sw and the lens finding robots and then attempt to understand why each
method failed to detect lenses from the other sample.

\subsection{Comparison of the \rf, \sw and \af samples}

\begin{figure}
\begin{center}
\includegraphics[scale=0.65]{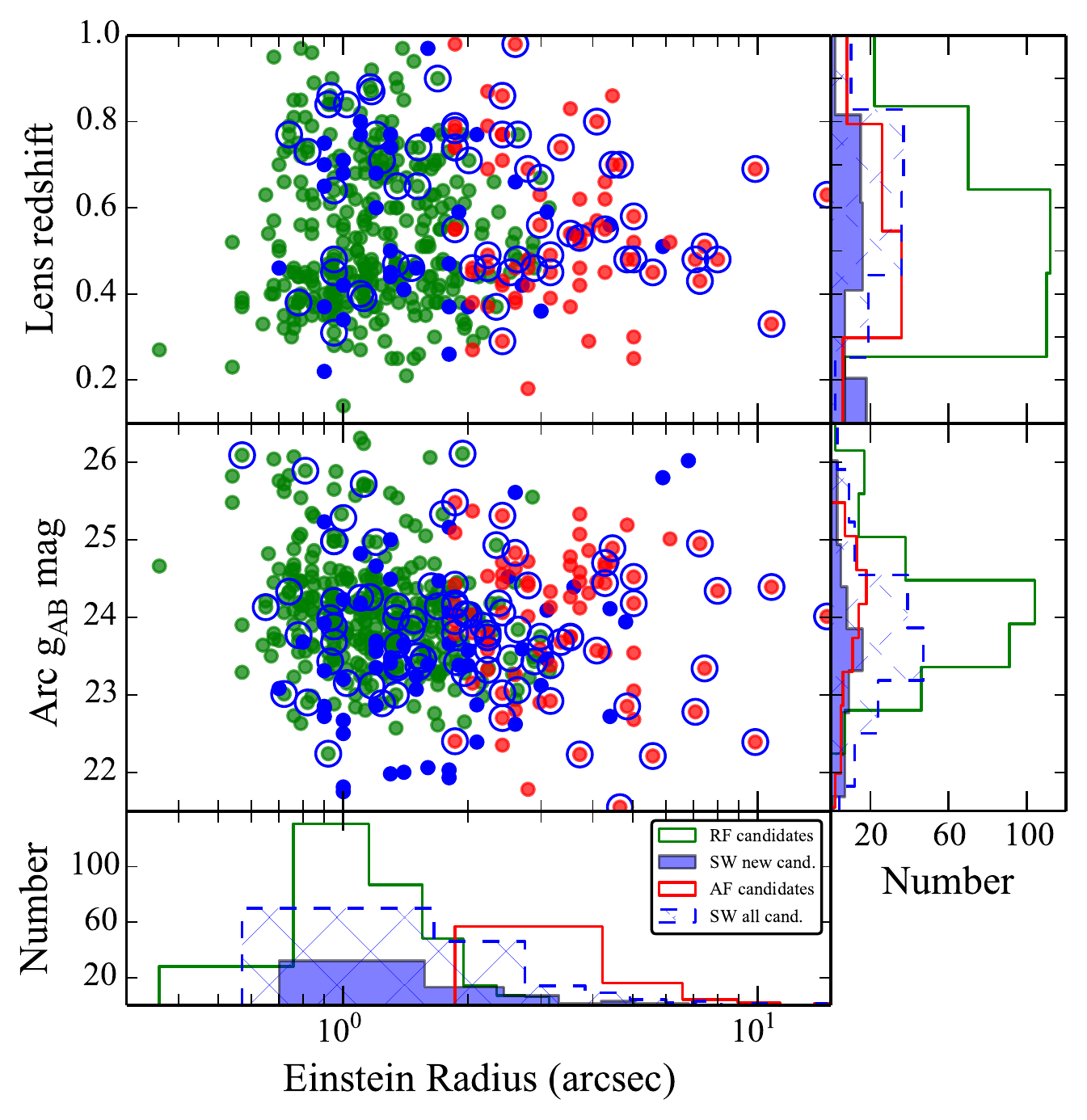}
\caption{ \label{fig:zlmgre}
Comparison of the lens redshift and the arc magnitude with the
Einstein radius for all of the three lens samples, namely, the \rf (green dots),
\sw (known candidates$-$blue circles and new candidates only$-$blue dots)
and \af (red dots). All samples have broadly similar properties.}
\end{center}
\end{figure}

In \Fref{fig:zlmgre}, we show the lens redshift and the arc flux
measured in $g$-band AB magnitude as a function of the Einstein radius for the \rf
(green), the \af (red) and the \sw sample (new candidates only in blue
and known candidates as blue circles). We note that the errors on the
redshift measurement should not be too different across the samples
since they are measured by a single method. However, the error on the
total flux of the lensed images are likely to be different across the
samples and the types of systematics are also different.  We have not
attempted to quantify these errors in this work. With that caveat, we
find that the \sw candidates sample is broadly similar to the
robotically found lens candidates in terms of the flux of the lensed
images and the redshift of the lensing galaxies.

%
%

The properties considered here do not show any clear differences between
the types of lenses being found by each method. Other properties such as
the flux of the lensing galaxies and the surface brightness of the
lensed images may be useful in showing some qualitative differences but
this is beyond the scope of our current analysis. A more detailed and
accurate analysis is deferred to the future.

\begin{figure*}
\begin{center}
\includegraphics[scale=0.5]{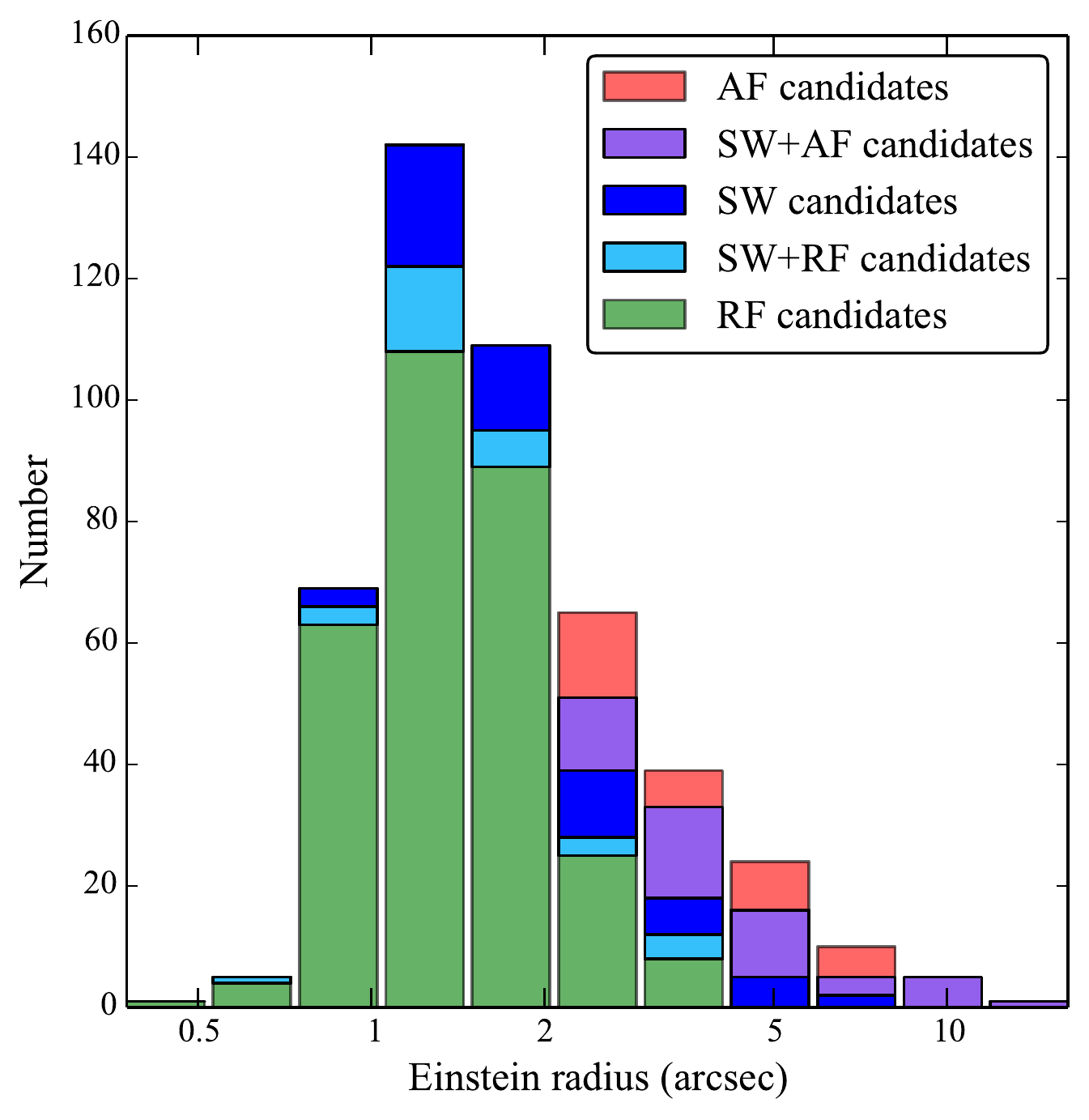}
\includegraphics[scale=0.5]{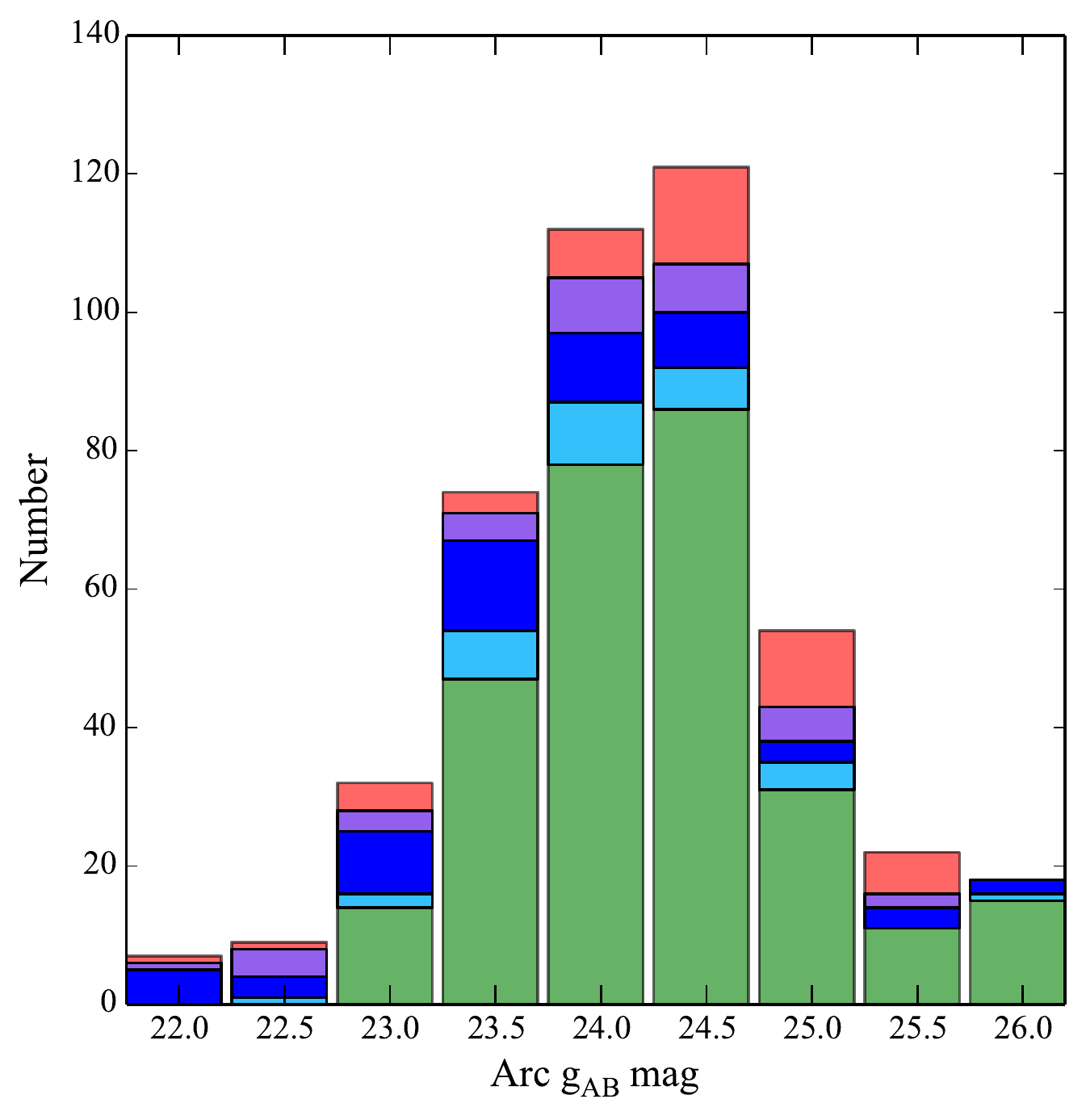}
\caption{ \label{fig:stackremg}
Candidate detections by the \rf, \sw and the \af as a
function of the Einstein radius and $g$ band magnitude of the lensed
images.
}
\end{center}
\end{figure*}

In \Fref{fig:stackremg}, we show the relative distribution of
number of candidates from each sample as a function of the Einstein
radius and arc magnitude. The light blue colour shows the overlap between
the \sw and the \rf samples and the purple colour shows the overlap
between the \sw candidates and the \af samples. As noted earlier, the
\rf dominates the small $R_{\rm E} (<2\arcsec)$ detections although \sw does
find a modest number of candidates in this range. At larger $R_{\rm E}$,
\sw sample begins to dominate and is comparable to the \af sample.  As a
function of the arc magnitudes, all three samples have detections at all
magnitudes and median magnitudes for all samples is around $g\sim24.5$.
Relatively, the \rf sample spans a narrower range compared to the \sw
and \af sample. However, this can be verified only after understanding
and accounting for the systematic uncertainties in our measurements.

\subsection{Why were the new \sw candidates missed by the robots?}

We test the \rf and \af on images centred on the new \sw candidates to
trace and understand at what stage the algorithm failed to detect them.

First, we re-ran \rf on the new \sw sample. At the beginning, a galaxy
catalog is generated based on magnitude, redshift and SED type
\citep[see]{Gavazzi2014} to select galaxies which are most likely to act
as lenses. We find that about 40~per~cent of the new \sw candidates failed to
meet this initial selection criteria, for example, SW1, SW14, SW20,
SW23, SW27 and SW30. All of the lensing galaxies are bright enough to
satisfy the magnitude criterion ($i<22$). However, some of them have a
bright companion galaxy, some of them do not look like E/S0 type
galaxies and some are edge on galaxies which could be the reason for
these galaxies having failed the photometric redshift and SED type pre-selection.

In subsequent \rf steps, the flux from the galaxy is subtracted from the
scaled difference image to enhance the visibility of the faint blue
lensed features. An object finder is then run on this image to quantify the
lensed image properties.

Another $\sim$~50~per~cent of the \sw candidates could not be
detected by the object finder because properties such as the image area,
axis ratio, magnitude/colour and alignment with respect to the lensing
galaxy were not satisfied. Some of the candidates missed at this stage
are, for example, SW4, SW5, SW6, SW26, SW36, SW39 and SW46.

Next, we re-ran the \af on the same \sw sample of new candidates. The \af
is directly run on the images to look for elongated arc like objects and does
not require a list of targets to begin with. Objects are identified by
placing thresholds on the noise level in the images. Thus, \af
detections are sensitive to changes in the noise levels.

Originally, the \af was run on a large image with an area of $\sim 19350
\times 19350$ pixels$^2$. For the rerun, we worked with much smaller
images because this is faster. However, this alters the measured
noise and hence affects the number and type of arc detections. We find
that about 30~per~cent of the new candidates were detected without changing any
of the thresholds in the code, suggesting that these could have been detected by
\af had its noise thresholds been set differently.

The \af code calculates second order brightness moments around every
pixel to decide if the distribution of flux is elongated in some
direction in order to detect elongated arc-like objects. An elongation
estimator is assigned to every pixel. All pixels with a value of the
elongation estimator above a certain threshold are connected to form the arc
feature. This is called the segmentation of the arc candidate. Subsequently, arc
properties such as the area, mean flux, length and curvature are
determined. We relaxed the threshold at the segmentation stage and also relaxed
thresholds mainly on the area of the arc. These new settings led to the detection of
about 75~per~cent of the new \sw candidates. We find that relaxing thresholds
on other arc properties does not improve the detection rate significantly.

Typically then, the \sw candidates were missed from the \af sample
either because a) the arcs were fainter, b) the flux of the arc and the
galaxy were blended together (such that the \af incorrectly connected
part of the galaxy to the arc), c) the arcs were unusually short or
thick, or d) the lensed images are almost circular or point-like (\af
was not designed to detect lensed quasars).

Relaxing the \af thresholds obviously increases the number of candidate arc
detections but this also increases the false positive rate.
For example, the number of arc candidate detections increased by a factor
of $\sim$2 when we relaxed the thresholds in the rerun described above,
while the number of false positives increased by a factor of $\sim$5.
While the \af sample purity could be increased by
cross-correlating the arc candidate positions with a putative lens
galaxy catalog, these numbers illustrate the predicament facing automated
lens-finding algorithms, and the continuing benefits of visual screening.


\subsection{False negatives: known lenses missed by \sw}
\label{sec:fn}

Like any lens finding method, the \sw system could potentially be
failing to detect certain kinds of lenses.  We find that about
35~per~cent of the known sample of lenses are missed at \StageOne; 
about 10~per~cent losses were incurred during the \StageTwo refinement
(see \Tref{tab:stats}).  Below, we focus on the known lens sample at
\StageOne to understand why some of them are being missed and possibly
find a way to improve the detection rate which can be adopted in the
future \sw lens searches.

Many of the missed lenses are from the \rf sample with small
Einstein radii and faint lensed images (see \Fref{fig:compre}). Among
the confirmed lenses from \rf, about 45~per~cent are missed. Out of the
missed sample of 15 lenses, about half of them are visually difficult to
detect. The other half appear to have faint blue smudges around
galaxies which should have been easier to identify. Similarly,
if we consider the \af lens sample, $\sim$20~per~cent are missed by \sw. This
is a relatively small sample of $\sim 5$ systems and visual inspection
of them suggests that, by and large, either the lensed features are faint
or they have odd properties which makes them difficult to identify
correctly. For further tests, we combine the \rf and \af sample.

For a lens finding method which uses the collective skill, experience and
knowledge of a group of volunteers, it may be difficult to find a single
factor with certainty which causes a lens candidate to be missed. We
attempt to understand whether there is indeed a single dominant factor
that is resulting in the loss of these lenses, or if the lenses are being
missed due to a combination of multiple reasons. Below, we consider some
of the factors that could affect the efficiency of finding lenses.

\subsubsection{Number of classifications}

First, we check whether the number of classifications (Nclass) is
significantly lower for the missed sample compared to the detected one.
Most of the lenses in the known sample (including both those that were
detected and those that were missed)
received similar numbers of classifications to the other subjects.
A few received Nclass$>$20, possibly as a result of
continuing to remain for a long time in the database because there was
uncertainty over whether or not to reject them.
Overall, we do not find any difference in the number of classifications
between the detected and the missed lenses.

\subsubsection{Lens positions within the image cutouts}

The efficiency of a visual search could potentially
vary in different sections of an
image. Our eyes tend to focus usually at the centre of an image and lens
candidates close to the borders could go undetected. Therefore, it is
important to check whether \sw could be missing some of the known lenses
because they happen to be close to the borders of the image cutouts.

\begin{figure*}
\begin{center}
\includegraphics[scale=0.95]{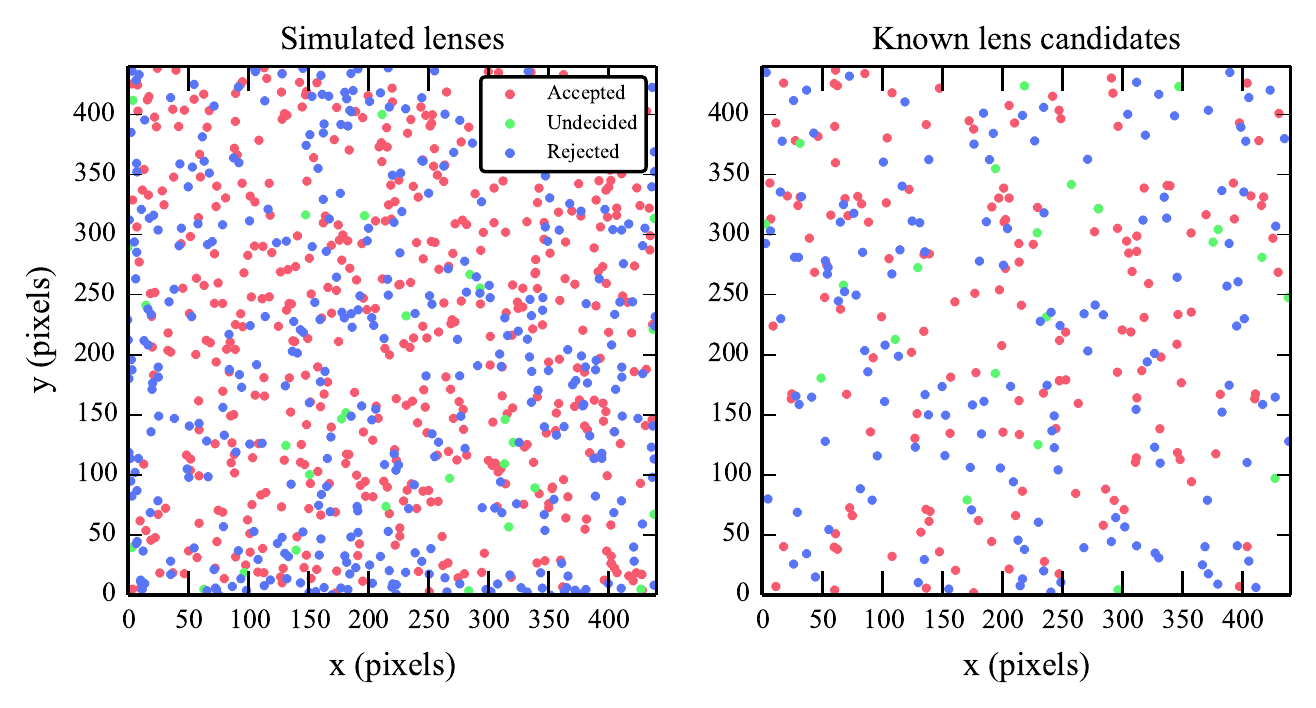}
\caption{ \label{fig:comppos}
Completeness as a function of the positions of the lens systems. Simulated lenses
(left) and real lens candidates (right) are shown. Irrespective of the
status of the lenses, that is, detected, undecided or rejected, there is
no strong dependency on the location of the lenses, both for the
simulated and the real sample of candidates. }
\end{center}
\end{figure*}

From the SWAP, the image cutouts inspected by the \sw volunteers receive
a status of detected (if $P>P_{\rm acc thresh}$), rejected (if
$P<P_{\rm rej thresh}$) and undecided (if $P_{\rm rej thresh}< P <
P_{\rm acc thresh}$). In \Fref{fig:comppos}, we compare the positions
of lenses which are detected (red), undecided (green) and rejected
(blue). The left and the right panels have the simulated lens sample and
the known lens candidates sample, respectively. We note that the density
of points do not represent the actual number of detections because, for
cases with large sample size, randomly drawn sub-samples are shown for
the ease of visual comparison.

We do not find any strong correlation in the detection rate of
lenses as a function of their positions in the image, for either the
simulated or the known lens sample. Thus, the completeness of the lens
sample is most likely not significantly affected by lenses
located close to the image borders.

\subsubsection{Classification Power}

Each \sw image classification is based on the markers placed by around 10
volunteers (on average, \PaperOne). This number could be
small enough to introduce some scatter in the system performance, arising from
the variations between the small groups of volunteers inspecting each subject.
In \PaperOne we investigated the system performance in terms of the ``Skill''
of each agent; in \Aref{appendix:power} we define a complementary property,
the ``Power'' of a classification to make a large difference in the probability
$P$ of an image containing a lens. Here, we investigate whether the distribution
of classification power is systematically different between the detected and missed lenses.

\begin{figure*}
\begin{center}
\includegraphics[scale=0.6]{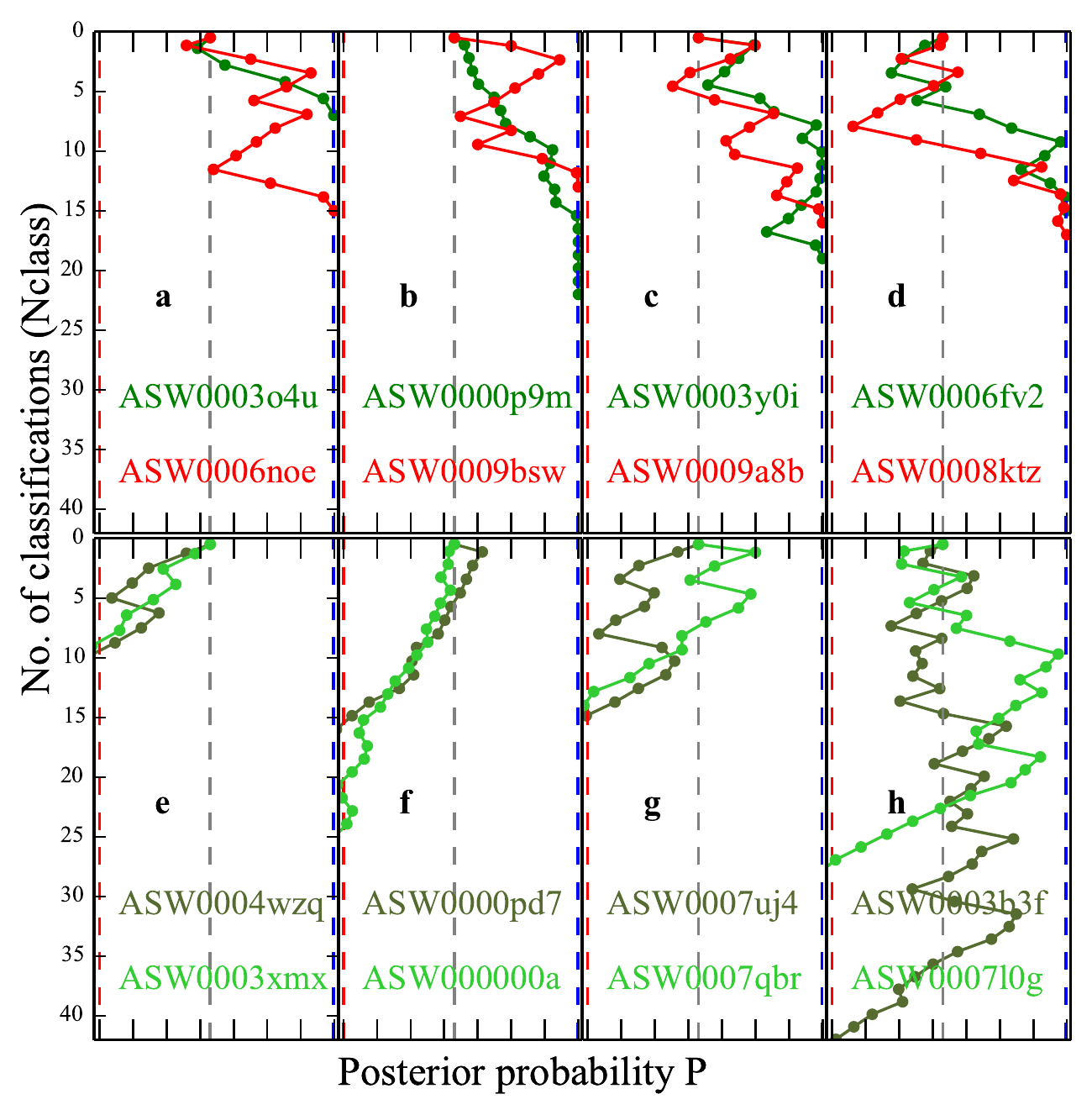}
\includegraphics[scale=0.6]{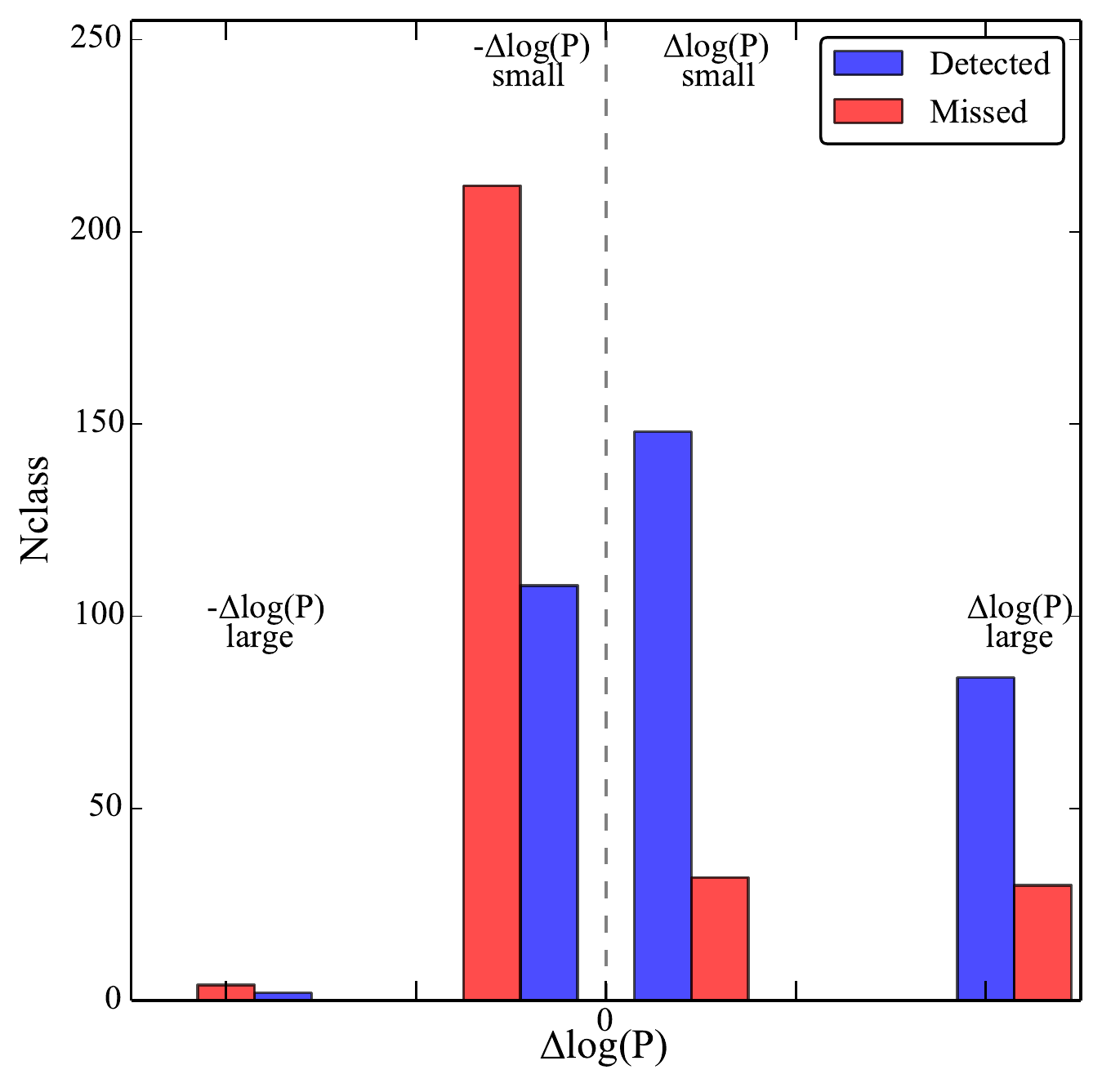}
\caption{ \label{fig:detmis}
Left: Examples of \StageOne trajectories of some known lenses.  Upper
panels show detected lenses, lower panels show missed lenses (false
negatives). The green trajectories for the detected sample are
counterparts of the light and dark green trajectories of the missed sample
except that the kicks are positive for the detected sample. The red
trajectories of the detected sample demonstrate that sufficient number
of positive large kicks can lead to the detection of the lenses. Within the
missed sample, the light and dark green correspond to visually easier and more
difficult to identify systems, respectively. Right: Histogram of classification
power received by known lenses.  Among the sample of missed lenses, most
classifications are from the low power volunteers identifying these
images incorrectly (i.e.  $-\Delta$~log(P)) which overshadows the small
number of correct identifications (i.e.  $\Delta$~log(P)) coming from
both the less and the high power volunteers combined. For the detected
lens sample, most classifications are correct identifications coming
from both low and high power volunteers. }
\end{center}
\end{figure*}

We check how the posterior probability $P$ (see \PaperOne for the
mathematical definition) of an image or a subject to contain a lens
changes as the image receives more classifications from multiple
volunteers. A graphical representation of changing probabilities for
increasing classifications is called a trajectory plot. In
\Fref{fig:detmis}, we show the trajectory plots of a few examples of
detected lenses (top row of panels) and missed lenses (bottom row of panels)
from \StageOne of \sw. Every subject is assigned a
prior probability $P_0=2\times10^{-4}$ (grey dashed line) and starts at
the middle of the trajectory plot. The number of classifications (Nclass) for a
subject increases from top to bottom (subjects move down the trajectory plots
as they are classified).  The $P$ value of a subject is updated
with every classification from the volunteer.  If a volunteer identifies
a lens candidate, the trajectory moves to the right otherwise moves to
the left. A subject is accepted if it crosses the blue-dashed line
marking the $P_{\rm acc thresh}$ (set to 0.95 for \StageOne) on the
right. It is rejected if it crosses the red-dashed line marking the
$P_{\rm rej thresh}$ (set to $10^{-7}$ for \StageOne) on the left.

The amount by which the posterior probability $P$ value of a subject will
change depends on how well the volunteers are performing on the training
sample and its current probability. Thus, for a given current
probability, some volunteers will change the $P$ by a large factor
compared to others. This is evident in the trajectory
plots, which show both large and small distances between consecutive points
which we refer to as kicks. Comparison of the kick sizes between the
detected and the missed lenses suggests that {\it the missed lenses do not
have as many volunteers giving large kicks}. We also note that most of
the large kicks seen in the trajectories of the missed lenses seem to be
moving the subjects to the right. In other words, certain ``high power''
volunteers are mostly classifying them as subjects with lens candidates.

The bottom panels of \Fref{fig:detmis}, show the trajectories of missed
lenses for the cases which are visually easier (light green) and more
difficult (dark green) to identify.  In spite of some mild
qualitative differences, both set of trajectories have very similar
behaviour. The trajectories in panel (e) are typical of this sample in
terms of Nclass and the dominance of small negative kicks. Panel (f)
represents a small fraction of this sample where the kicks are only
small and negative. The panel (g) shows how some lenses receive a bunch
of large positive kicks which are led to rejection by still mostly small
negative kicks. Finally, panel (h) shows those cases of lenses which
received almost sufficient number of large positive kicks to be detected
but ended up being rejected.

The detected lenses shown with green trajectories, in the upper
panels of \Fref{fig:detmis}, can be thought of as counterparts of the trajectories of the
missed lenses in the corresponding bottom panels except that their
classification power is different. Most detected lenses are similar to the case
in panel (a) that are detected within a few classifications coming from
large positive kicks.  Panel (b) represents a few odd cases which are
dominated mainly by small positive kicks. Panel (c) shows a lens getting
more classifications, but not reaching the detection or rejection
thresholds because of the tug between positive and negative kicks mostly
from experienced volunteers.  Panel (d) represents two extreme cases when
the images are on the verge of being rejected but are saved thanks to a
series of large positive kicks.  The red trajectories are some more
examples of randomly selected cases which demonstrate how having
sufficient number of large positive kicks allows lenses to be detected
in spite of several small negative kicks.

For a quantitative comparison of the large and small kicks for the
entire samples of detected and missed known lenses, we show a plot of
histogram on the right of \Fref{fig:detmis}. Qualitatively, there are
four types of volunteers making classifications: those causing large, positive
kicks (correct classifications by high power volunteers), those
causing small, positive kicks (correct classifications by low power
volunteers), those causing small, negative kicks (incorrect
classifications by low power volunteers) and those causing large, negative
kicks (incorrect classifications by high power volunteers). The
four histograms in the figure correspond to these four types of
volunteers for each sample (that is, detected or missed). In this plot,
the kick size
is defined as small if
$\Delta \log{P} (= \log{P_{\rm current}} - \log{P_{\rm
previous}}) < \Delta\log{P}_{\rm cut}$ (chosen as 1.2) and is large
if greater than $\Delta\log{P}_{\rm cut}$.

Some of the key inferences are as follows. i) The ratio of positive
kicks to negative kicks for the detected sample is higher than in the
missed sample, suggesting that the fraction of volunteers making
positive kicks is higher for the detected sample. ii) The number of
classifications received by the missed sample is dominated by small
negative kicks. In contrast, for the detected sample there are
comparable contributions from all three types -- small positive kicks,
large positive kicks, and small negative kicks.  iii) The number of
classifications providing large negative kicks is lower for both the
detected and the missed samples. This is consistent with our expectation
that high power volunteers should not be making incorrect
classifications.  We conclude that one of the major factors in the \SW
system missing the known lenses is a lack of high power classifications.

As a demonstration, we re-ran SWAP for \StageOne using only  classifications
that produced large kicks, with $|\Delta \log{P}| > 1.2$.  This obviously meant
reducing the total number of classifications per subject by a large fraction.
As a result, we also needed to change the $P_{\rm acc thresh}$. Choosing this to
be 0.1, we found that about a third of the lenses that had previously been
missed were now detected, while all the previously detected lenses were again
detected. The remaining missed lenses simply do not have enough classifications
from volunteers producing large positive kicks. This experiment shows that it
may be possible to increase the  \sw completeness  by preferentially showing
certain rejected systems -- those that  had never received a high power
classification --  to  volunteers capable of making such classifications.

Changing the rejection and acceptance thresholds will likely 
decrease the purity along with improved completeness. This will
need to be further quantified before detailed recommendations can be
made. However, dynamically assigning certain subjects to volunteers
according to various measures of their skill seems like a fruitful line
of investigation when seeking to improve the system performance in
future.

\section{Summary and Conclusions}
\label{sec:conclude}

We report the discovery of gravitational lens candidates from the
first \sw lens search. In this search, volunteers were shown $g-r-i$
colour images of random regions of the sky taken by the CFHT Legacy
Survey. The aim of this blind lens search was to find lenses that had
been missed by previous searches done on the \cfhtls with lens finding
algorithms.

The search was carried out in two stages.  In \StageOne, volunteers
inspected $\sim$ 430,000 images, and selected a smaller sample of $\sim$
3000 images as having interesting lens candidates. In \StageTwo, after a
careful second inspection of the candidates from \StageOne, a purer
sample of $\sim$~500 candidates was obtained. In a final step, these
images were inspected by three of us (AM, AV and PJM) to produce a
sample of candidates with grades ranging from possibly a lens (1) to
almost certainly a lens (3).  In this paper, we presented this new \sw
sample and compared it with the previously known samples from two
robotic searches from the \cfhtls, namely, \rf and \af.

Our conclusions are as follows:
\begin{itemize}
\item \sw works well as a discovery engine for gravitational lenses
through citizen science. While a targeted visual search may be more
efficient, we show that the blind search works reasonably well too.

\item We use a sample of simulated lenses, duds and impostors tailored
to the \cfhtls data to train the volunteers and calibrate their
performance. The volunteers not only perform well on the training sample
(see Paper I) but also find lenses that are fainter, more compact or
redder than covered by the training sample demonstrating their
adaptability in this task. 

\item We present a sample of {\bf 29 new} gravitational lens candidates,
and an additional 30 medium grade systems. These 59 candidates received
averaged grade G$\ge$1.3 from three experts following the scale where
(1) means possibly, (2) means probably, and (3) means almost certainly,
a lens. In addition, among the G$\ge$1.3 sample, we re-discovered 82
lens candidates from various samples published in the literature.

\item Compared to the sample of \rf and \af robotically detected lens
candidates, the \sw sample finds lens systems with statistically similar
properties, including the range of lens redshifts, lensed image total
magnitudes, and Einstein radii. However, having only displayed images
without the lens galaxy light subtracted, the \sw sample does not
contain many of the \rf-identified-lensed images with sub-arcsecond
$R_{\rm E}$, just because the flux of the typically faint lensed images
is obscured by the flux from bright lensing galaxies.

\item Qualitatively, \sw seems to have found lens systems with different
types of lensing galaxies, for example, elliptical, spiral (face on and
edge on) and small red galaxies unlike those found from robotic
searches. Similarly, the lensed images too have diverse properties such
as different colours, morphologies and sizes which are again typically
missed by any given algorithm.

\item Based on the known sample of lenses and lens candidates, we find
that we lose a small fraction of them during \StageTwo refinement.  It
is more important to improve the lens detection sensitivity at the
initial \StageOne classification step.  About 35\% of the known lenses
(20 in total) were missed at \StageOne.  Two thirds of these missed
lenses were found to be galaxy-scale \rf systems, with faint arcs
blended with the bright lens galaxies.

\item It is possible to improve the \sw completeness by
changing the strategy of when and who is shown an image: only using high power
classifications recovers 40\% of the missed lenses.

\end{itemize}

The discovery of many new lens candidates through the first \sw lens
search has demonstrated that the citizen scientists have successfully
taught themselves to identify lenses within a short span of time.  They
have found lens candidates which the algorithms failed to discover.
Upcoming and planned wide-field imaging surveys such as the DES, HSC,
Euclid and LSST will produce formidable amounts of data. Blind lens searches as
described here will be impractical with these very large surveys. However,
it should be possible to conduct a blind search on a sub-area of a large survey in
order to assess the performance of either the algorithms or the
volunteers, the results of which can be extrapolated to the entire survey.  As
demonstrated in this paper, any one approach for finding lenses from the
entire survey data may not be sufficiently complete and pure. Thus, combining
robotic methods for pre-selection with the citizen science approach for
visual screening might be a good strategy for finding lenses in these
large imaging surveys.

\onecolumn
\begin{center}
\begin{longtable}{llrrrrrrlrr}
\caption{ \label{tab:swcands}
Sample of the \sw new lens candidates. }\\
\hline
SW ID & Name & RA & Dec &  $z_{\rm phot}$ & $m_i$ & $R_{\rm E}$ & $G$ & ZooID & $P$ & Comments  \\
  &  & (deg) & (deg) &  & (mag) &  (") &  &  & & \\
\hline
\endfirsthead
\hline
SW ID & Name & RA & Dec &  $z_{\rm phot}$ & $m_i$ & $R_{\rm E}$ & $G$ & ZooID & $P$ & Comments  \\
  &  & (deg) & (deg) &  & (mag) &  (") &  &  & & \\
\hline
\endhead
\hline
\multicolumn{11}{p{18cm}}{
The column Comments has two type of notes. The first is about the lens
image configuration where the symbols mean the following A: Arc, D:
Double, Q: Quad, R: Ring. The second is a comment on the type of lens
assessed visually. Note that this classification is not based on colours
or spectral analysis. The symbols are E: Elliptical, S: (face on)
Spiral, G: Group-scale, D: Edge on disk, R: Red star-forming galaxy.
}\\
\endlastfoot
SW19 & CFHTLS\,J020642.0$-$095157 &  31.67504 &  $-$9.86584 &  0.2 & 20.8 &  0.9 &  2.0 & ASW0001ld7 &  0.8 &  A,R   \\ 
SW8  & CFHTLS\,J020648.0$-$065639 &  31.70031 &  $-$6.94430 &  0.8 & 20.2 &  1.3 &  2.3 & ASW00099ed &  0.4 &  A,E   \\ 
SW43 & CFHTLS\,J020810.7$-$040220 &  32.04497 &  $-$4.03891 &  1.0 & 20.8 &  1.8 &  1.3 & ASW0001c3j &  0.7 &  A,R   \\ 
SW9  & CFHTLS\,J020832.1$-$043315 &  32.13396 &  $-$4.55429 &  1.0 & 21.0 &  1.6 &  2.3 & ASW0002asp &  1.0 &  A,R   \\ 
SW10 & CFHTLS\,J020848.2$-$042427 &  32.20110 &  $-$4.40751 &  0.8 & 20.5 &  1.1 &  2.3 & ASW0002bmc &  0.9 &  D,D   \\ 
SW11 & CFHTLS\,J020849.8$-$050429 &  32.20784 &  $-$5.07494 &  0.8 & 20.6 &  0.9 &  2.3 & ASW0002qtn &  1.0 &  A,R   \\ 
SW44 & CFHTLS\,J021021.5$-$093415 &  32.58981 &  $-$9.57109 &  0.4 & 18.4 &  2.7 &  1.3 & ASW0002k40 &  0.4 &  D,S   \\ 
SW30 & CFHTLS\,J021057.9$-$084450 &  32.74148 &  $-$8.74745 &  0.0 &  0.0 &  2.5 &  1.7 & ASW0002p8y &  0.4 &  A,G   \\ 
SW20 & CFHTLS\,J021221.1$-$105251 &  33.08810 & $-$10.88106 &  0.3 & 17.9 &  1.8 &  2.0 & ASW0002dx7 &  0.8 &  D,E/S   \\ 
SW45 & CFHTLS\,J021225.2$-$085211 &  33.10511 &  $-$8.86973 &  0.8 & 19.5 &  2.1 &  1.3 & ASW00024id &  1.0 &  R,R   \\ 
SW46 & CFHTLS\,J021317.6$-$084819 &  33.32341 &  $-$8.80549 &  0.5 & 19.8 &  1.3 &  1.3 & ASW00024q6 &  0.4 &  A,R/E   \\ 
SW31 & CFHTLS\,J021514.6$-$092440 &  33.81089 &  $-$9.41115 &  0.7 & 19.9 &  2.6 &  1.7 & ASW00021r0 &  0.4 &  A,R/G   \\ 
SW32 & CFHTLS\,J022359.8$-$083651 &  35.99955 &  $-$8.61439 &  0.0 &  0.0 &  3.1 &  1.7 & ASW0004iye &  0.4 &  A,E   \\ 
SW12 & CFHTLS\,J022406.1$-$062846 &  36.02558 &  $-$6.47963 &  0.4 & 19.6 &  0.9 &  2.3 & ASW0003wsu &  0.7 &  A,E   \\ 
SW1  & CFHTLS\,J022409.5$-$105807 &  36.03978 & $-$10.96885 &  0.0 &  0.0 &  4.8 &  3.0 & ASW0004dv8 &  1.0 &  A,G   \\ 
SW21 & CFHTLS\,J022533.3$-$053204 &  36.38882 &  $-$5.53460 &  0.5 & 19.4 &  3.6 &  2.0 & ASW0004m3x &  0.4 &  A,R/G   \\ 
SW22 & CFHTLS\,J022716.4$-$105602 &  36.81856 & $-$10.93410 &  0.4 & 17.3 &  1.8 &  2.0 & ASW0009ab8 &  0.7 &  A,E/G   \\ 
SW33 & CFHTLS\,J022745.2$-$062518 &  36.93868 &  $-$6.42183 &  0.6 & 20.5 &  1.2 &  1.7 & ASW0003s0m &  0.5 &  A,R   \\ 
SW13 & CFHTLS\,J022805.6$-$051733 &  37.02362 &  $-$5.29266 &  0.4 & 18.8 &  1.4 &  2.3 & ASW00047ae &  1.0 &  Q,E   \\ 
SW47 & CFHTLS\,J022843.0$-$063316 &  37.17942 &  $-$6.55465 &  0.5 & 19.1 &  1.8 &  1.3 & ASW0003r6c &  0.3 &  D/A,E   \\ 
SW23 & CFHTLS\,J023008.6$-$054038 &  37.53591 &  $-$5.67744 &  0.6 & 19.7 &  1.9 &  2.0 & ASW0003r61 &  0.5 &  A,E   \\ 
SW14 & CFHTLS\,J023123.2$-$082535 &  37.84682 &  $-$8.42663 &  0.0 &  0.0 &  1.2 &  2.3 & ASW0004xjk &  0.3 &  A,R   \\ 
SW24 & CFHTLS\,J023315.2$-$042243 &  38.31334 &  $-$4.37886 &  0.7 & 19.7 &  1.0 &  2.0 & ASW00050sk &  0.8 &  A,R   \\ 
SW34 & CFHTLS\,J023453.5$-$093032 &  38.72321 &  $-$9.50892 &  0.5 & 19.8 &  0.7 &  1.7 & ASW00051ld &  0.3 &  A,D   \\ 
SW35 & CFHTLS\,J084833.2$-$044051 & 132.13847 &  $-$4.68085 &  0.8 & 20.2 &  0.9 &  1.7 & ASW0004wgd &  0.7 &  A,R   \\ 
SW15 & CFHTLS\,J084841.0$-$045237 & 132.17084 &  $-$4.87720 &  0.3 & 19.0 &  1.0 &  2.3 & ASW0004nan &  1.0 &  A,E   \\ 
SW48 & CFHTLS\,J090219.0$-$053923 & 135.57947 &  $-$5.65666 &  0.0 &  0.0 &  2.0 &  1.3 & ASW0000g95 &  1.0 &  A,R/E   \\ 
SW36 & CFHTLS\,J090248.4$-$010232 & 135.70204 &  $-$1.04243 &  0.4 & 19.1 &  1.4 &  1.7 & ASW000096t &  0.6 &  D,E   \\ 
SW25 & CFHTLS\,J090308.2$-$043252 & 135.78449 &  $-$4.54789 &  0.0 &  0.0 &  1.3 &  2.0 & ASW00007mq &  0.6 &  D,D   \\ 
SW49 & CFHTLS\,J090319.4$-$040146 & 135.83105 &  $-$4.02971 &  0.0 & 19.8 &  1.2 &  1.3 & ASW00007ls &  0.5 &  A,R/E   \\ 
SW50 & CFHTLS\,J090333.2$-$005829 & 135.88869 &  $-$0.97490 &  0.0 &  0.0 &  2.1 &  1.3 & ASW00008a0 &  1.0 &  A/D,E/G   \\ 
SW51 & CFHTLS\,J135724.8$+$561614 & 209.35374 &    56.27066 &  0.0 &  0.0 &  2.6 &  1.3 & ASW0006e0o &  0.9 &  D,E   \\ 
SW26 & CFHTLS\,J135755.8$+$571722 & 209.48268 &    57.28971 &  0.8 & 20.2 &  0.9 &  2.0 & ASW0005ma2 &  0.8 &  A,R   \\ 
SW52 & CFHTLS\,J140027.9$+$541028 & 210.11636 &    54.17455 &  0.0 &  0.0 &  1.2 &  1.3 & ASW0006a07 &  0.6 &  Q,R/E   \\ 
SW16 & CFHTLS\,J140030.2$+$574437 & 210.12601 &    57.74371 &  0.4 & 18.2 &  2.0 &  2.3 & ASW0009bp2 &  0.6 &  A,E   \\ 
SW2  & CFHTLS\,J140522.2$+$574333 & 211.34261 &    57.72587 &  0.7 & 19.7 &  1.0 &  2.7 & ASW000619d &  0.7 &  A,R   \\ 
SW17 & CFHTLS\,J140622.9$+$520942 & 211.59581 &    52.16169 &  0.7 & 20.3 &  1.2 &  2.3 & ASW0005rnb &  0.7 &  A,R   \\ 
SW27 & CFHTLS\,J141432.9$+$534004 & 213.63716 &    53.66788 &  0.7 & 21.4 &  1.0 &  2.0 & ASW0006jh5 &  0.8 &  A,R   \\ 
SW53 & CFHTLS\,J141518.9$+$513915 & 213.82903 &    51.65420 &  0.4 & 18.3 &  3.0 &  1.3 & ASW00070vl &  0.8 &  D,E   \\ 
SW3  & CFHTLS\,J142603.2$+$511421 & 216.51375 &    51.23935 &  0.0 &  0.0 &  4.4 &  2.7 & ASW0006mea &  0.7 &  A,G   \\ 
SW54 & CFHTLS\,J142620.8$+$561356 & 216.58699 &    56.23230 &  0.5 & 19.5 &  1.3 &  1.3 & ASW0007sez &  0.8 &  A/R,S   \\ 
SW55 & CFHTLS\,J142652.8$+$560001 & 216.72004 &    56.00044 &  0.0 &  0.0 &  1.5 &  1.3 & ASW0007t5y &  1.0 &  R,R   \\ 
SW56 & CFHTLS\,J142843.5$+$543713 & 217.18153 &    54.62036 &  0.4 & 19.7 &  1.3 &  1.3 & ASW0007pga &  0.6 &  D,D   \\ 
SW4  & CFHTLS\,J142934.2$+$562541 & 217.39261 &    56.42807 &  0.5 & 19.0 &  5.9 &  2.7 & ASW0009cjs &  0.8 &  A,G   \\ 
SW28 & CFHTLS\,J143055.9$+$572431 & 217.73332 &    57.40883 &  0.7 & 19.3 &  1.6 &  2.0 & ASW0007xrs &  0.9 &  A,R/G   \\ 
SW37 & CFHTLS\,J143100.2$+$564603 & 217.75124 &    56.76750 &  0.0 &  0.0 &  1.2 &  1.7 & ASW00086xq &  0.8 &  A,E   \\ 
SW38 & CFHTLS\,J143353.6$+$542310 & 218.47357 &    54.38624 &  0.8 & 19.8 &  1.8 &  1.7 & ASW0009cp0 &  0.7 &  A,E   \\ 
SW5  & CFHTLS\,J143454.4$+$522850 & 218.72702 &    52.48080 &  0.6 & 19.4 &  4.4 &  2.7 & ASW0007k4r &  0.4 &  Q,G/R   \\ 
SW6  & CFHTLS\,J143627.9$+$563832 & 219.11636 &    56.64249 &  0.5 & 19.5 &  1.5 &  2.7 & ASW0008swn &  0.9 &  A,D   \\ 
SW57 & CFHTLS\,J143631.5$+$571131 & 219.13155 &    57.19215 &  0.7 & 20.9 &  1.3 &  1.3 & ASW0008pag &  0.6 &  D/A,R   \\ 
SW58 & CFHTLS\,J143651.6$+$530705 & 219.21503 &    53.11832 &  0.6 & 19.2 &  3.1 &  1.3 & ASW0007iwp &  0.7 &  A,E/G   \\ 
SW18 & CFHTLS\,J143658.1$+$533807 & 219.24246 &    53.63550 &  0.7 & 19.6 &  0.9 &  2.3 & ASW0007hu2 &  0.6 &  D,D   \\ 
SW29 & CFHTLS\,J143838.1$+$572647 & 219.65887 &    57.44645 &  0.8 & 20.2 &  1.1 &  2.0 & ASW0008qsm &  0.9 &  A,R   \\ 
SW59 & CFHTLS\,J143950.6$+$544606 & 219.96101 &    54.76858 &  0.0 &  0.0 &  1.7 &  1.3 & ASW00085cp &  0.4 &  A,G/R   \\ 
SW39 & CFHTLS\,J220215.2$+$012124 & 330.56348 &     1.35667 &  0.3 & 17.4 &  4.6 &  1.7 & ASW0005qiz &  0.5 &  rA,G   \\ 
SW7  & CFHTLS\,J220256.8$+$023432 & 330.73691 &     2.57581 &  0.0 &  0.0 &  6.8 &  2.7 & ASW0007e08 &  0.8 &  A,G   \\ 
SW40 & CFHTLS\,J221306.1$+$014708 & 333.27579 &     1.78561 &  0.0 & 17.1 &  1.4 &  1.7 & ASW0008wmr &  0.9 &  A,S   \\ 
SW41 & CFHTLS\,J221519.7$+$005758 & 333.83212 &     0.96615 &  0.4 & 20.2 &  1.0 &  1.7 & ASW0008xbu &  0.8 &  A,D   \\ 
SW42 & CFHTLS\,J221716.5$+$015826 & 334.31894 &     1.97394 &  0.1 & 21.6 &  1.0 &  1.7 & ASW00096rm &  1.0 &  A/R,R   \\ 
\end{longtable}
\end{center}

\twocolumn
\section*{Acknowledgements}

We thank all 36,982 members of the \sw community for their
contributions to the project so far. A complete list of registered
collaborators is provided at \texttt{http://spacewarps.org/\#/projects/CFHTLS}.
We also thank the anonymous referee for useful comments on the paper.

PJM was given support by the Royal Society, in the form of a research
fellowship, and by the U.S. Department of Energy under contract number DE-AC02-76SF00515.
AV acknowledges support from the Leverhulme Trust in the form of a research
fellowship.
The work of AM and SM was supported by World Premier International Research
Center Initiative (WPI Initiative), MEXT, Japan. AM acknowledges the support of
the Japan Society for Promotion of Science (JSPS) fellowship. The work of AM
was also supported in part by National Science Foundation Grant No.
PHYS-1066293 and the hospitality of the Aspen Center for Physics. 
%
%

The \sw project is open source.
The web app was developed at \texttt{https://github.com/Zooniverse/Lens-Zoo}, and was supported by a grant from the Alfred P. Sloan Foundation, 
while the SWAP analysis software was developed at
\texttt{https://github.com/drphilmarshall/SpaceWarps}.

The CFHTLS data used in this work are based on observations obtained with
MegaPrime/MegaCam, a joint project of CFHT and CEA/IRFU, at the
Canada-France-Hawaii Telescope (CFHT) which is operated by the National Research
Council (NRC) of Canada, the Institut National des Science de l'Univers of the
Centre National de la Recherche Scientifique (CNRS) of France, and the
University of Hawaii. This work is based in part on data products produced at
Terapix available at the Canadian Astronomy Data Centre as part of the
Canada-France-Hawaii Telescope Legacy Survey, a collaborative project of NRC and
CNRS.


\appendix

\section{Lens Detection Power}
\label{appendix:power}

\begin{figure*}
\begin{center}
\includegraphics[scale=1.0]{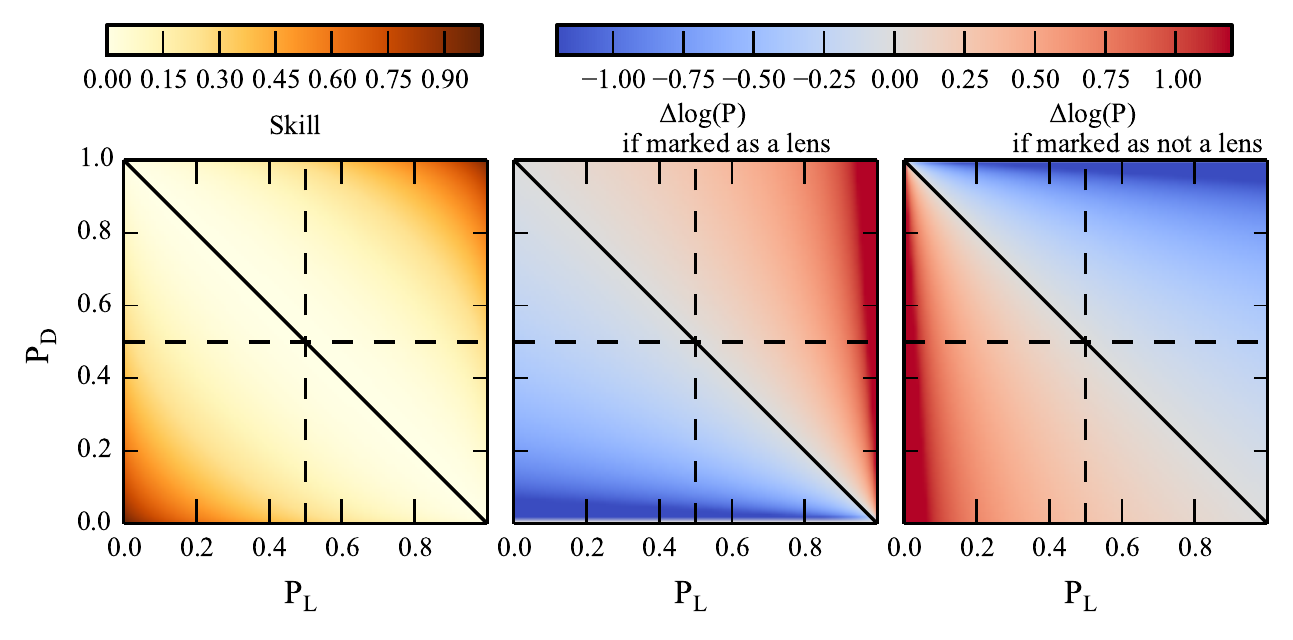}
\caption{ \label{fig:skilldlnp}
Skill of the volunteers and $\Delta\log{P}$ given a lens or not a lens in
an image as a function of $P_{\rm L}$ and $P_{\rm D}$. These
quantities indicate the ability of the volunteers but do not have a simple
linear relation with $P_{\rm L}$ and $P_{\rm D}$.
}
\end{center}
\end{figure*}

In \PaperOne, we defined the ``Skill'' of an agent as being given by the
expectation value of the information gain per classification. This
quantity is a non-linear function of both the $P_{\rm L}$, the
probability of correctly identifying a lens as a lens and $P_{\rm D}$,
the probability of correctly identifying a dud as a dud. This means that
one can get the same value of Skill for different combinations of
$P_{\rm L}$ and $P_{\rm D}$ (see the left panel of
\Fref{fig:skilldlnp}). The skill reflects the all-round ability of a
classifier to contribute information.

As described in \PaperOne, the posterior probability $P$
of a subject is determined by the $P_{\rm L}$ and $P_{\rm D}$ of all the
volunteers who clicked on the subject, via Bayes' Theorem. Each agent
will apply a ``kick'' of a different size to the subject probability,
$\Delta\log{P}$, which can be either positive (if the classifier thinks
the subject contains a lens) or negative (if the classifier thinks the
subject does not contain a lens). For instance, given a subject
containing a lens, a volunteer with high $P_{\rm L}$
implies a large positive kick irrespective of the value of
$P_{\rm D}$, as shown in the middle panel of
\Fref{fig:skilldlnp}. However, large positive kicks are
still possible for a volunteer located in the upper triangle
with different combinations of ($P_{\rm L}$, $P_{\rm D}$) suggesting that
the kick is not a simple function of ($P_{\rm L}$, $P_{\rm D}$).

The kicks appear as steps on the subject's trajectory plot. This kick
magnitude gives a useful measure of an agent's ``Power'' to move
images closer to detection.  Note that a volunteer who is very good at
rejecting duds, but not so good at identifying lenses, may have a high
Skill but a low Power (since they may fail to detect many of the
interesting lenses): Power provides a more precise quantification of a
classifier's ability to detect lenses (compared to rejecting non-lenses).

\bibliographystyle{apj}
\bibliography{references_cfhtls}


\label{lastpage}
\bsp

\end{document}